\documentclass[12pt]{article}
\usepackage{amsmath, amsfonts,amssymb}
\usepackage{graphicx}
\usepackage{bbm}
\usepackage{color}
\newcommand{\eq}[1]{\begin{equation}\label{#1}}
\newcommand{\en}{\end{equation}}
\newcommand{\ear}[1]{\begin{eqnarray}\label{#1}}
\newcommand{\enar}{\end{eqnarray}}

\newcommand{\SU}{\mathrm{SU}}
\newcommand{\plb}[3]{Phys. Lett. {\bf B #1}, #3 (#2)}
\newcommand{\prl}[3]{Phys. Rev. Lett. {\bf #1}, #3 (#2)}
\newcommand{\npb}[3]{Nucl. Phys. {\bf B #1}, #3 (#2)}
\newcommand{\prd}[3]{Phys. Rev. {\bf D #1}, #3 (#2)}
\newcommand{\jphg}[3]{J. Phys. {\bf G #1}, #3 (#2)}
\newcommand{\jhep}[3]{J. High Energy Phys. {\bf #1}, #3 (#2)}
\newcommand{\npps}[3]{Nucl. Phys. Proc. Suppl. {\bf #1}, #3 (#2)}

\usepackage{xspace}
\usepackage{epsfig}
\usepackage{setspace}

\newcommand{\mytitle}[1]{
                         \begin{center}
                           \LARGE{\textbf{#1}}
                         \end{center}}
\newcommand{\myauthor}[1]{\textbf{#1}}
\newcommand{\myaddress}[1]{\textit{#1}}
\newcommand{\mypreprint}[1]{\begin{flushright} #1 \end{flushright}}
\newcommand{\be}{\begin{equation}}
\newcommand{\ee}{\end{equation}}
\newcommand{\ba}{\begin{eqnarray}}
\newcommand{\ea}{\end{eqnarray}}

\begin{document}

\begin{titlepage}
\mypreprint{DESY 09-029}

\vspace*{0.5cm}
\mytitle{Semileptonic form factors $D\to\pi,K$ and $B\to\pi,K$ from a fine lattice}
\vspace*{0.3cm}

\begin{center}

\myauthor{A.~Al-Haydari,$^a$} \;
\myauthor{A.~Ali~Khan,$^a$} \;
\myauthor{V.~M.~Braun,$^b$} \;
\myauthor{S.~Collins,$^b$} \;
\myauthor{M.~G\"ockeler,$^b$} \;
\myauthor{G.~N.~Lacagnina,$^c$} \;
\myauthor{M.~Panero,$^{b, d}$} \;
\myauthor{A.~Sch\"afer$^b$} \;
{\bf  and }
\myauthor{G.~Schierholz$^{b, e}$}

  \vspace*{0.5cm}
\myaddress{$^a$
 Department of Physics, Faculty of Science, Taiz University, Taiz, Yemen}\\[2ex]
\myaddress{$^b$
 Institute for Theoretical Physics, University of
Regensburg, 93040 Regensburg, Germany}\\[2ex]
\myaddress{$^c$
 INFN, Sezione di Milano, 20133 Milano, Italy}\\[2ex]
\myaddress{$^d$
 Institute for Theoretical Physics, ETH Z\"urich, 8093 Z\"urich, Switzerland}\\[2ex]
\myaddress{$^e$
 Deutsches Elektronen-Synchrotron DESY, 22603 Hamburg,
  Germany}\\[2ex]

\vspace*{0.5cm}

QCDSF Collaboration

\end{center}

\vspace*{0.5cm}

\begin{abstract}

  \noindent We extract the form factors relevant for semileptonic
  decays of $D$ and $B$ mesons from a relativistic computation
  on a fine lattice in the quenched approximation. The lattice
  spacing is $a=0.04$~fm (corresponding to $a^{-1}=4.97$~GeV), which
  allows us to run very close to the physical $B$ meson mass, and to
  reduce the systematic errors associated with the
  extrapolation in terms of a heavy quark expansion. For decays of $D$
  and $D_s$ mesons, our results for the physical form factors at
  $q^2=0$ are as follows: $ f_+^{D\rightarrow\pi}(0)= 0.74(6)(4)$,
  $f_+^{D\rightarrow K}(0)= 0.78(5)(4)$ and $ f_+^{D_s\rightarrow K}(0)=
  0.68(4)(3)$. Similarly, for $B$ and $B_s$ we find: $
  f_+^{B\rightarrow\pi}(0)=0.27(7)(5)$, $f_+^{B\rightarrow K}(0)=
  0.32(6)(6)$ and $ f_+^{B_s\rightarrow K}(0)=0.23(5)(4)$. We compare
  our results with other quenched and unquenched lattice calculations,
  as well as with light-cone sum rule predictions, finding good
  agreement.
\end{abstract}
\vspace*{0.2cm}
\end{titlepage}

\section{Introduction}
\label{sec:intro}

Heavy meson decays are the main source of precision information on quark flavor mixing parameters in the Standard Model. The over-determination of the sides and the angles of the CKM unitarity triangle is the aim of an extensive experimental study: It addresses the question whether there is New Physics in flavor-changing processes and where it manifests itself. One of the sides of the unitarity triangle is given by the ratio $|V_{ub}/V_{cb}|$. $V_{cb}$ is known to approximately $2\%$ accuracy from $b\to c \ell \nu_\ell$ transitions~\cite{Amsler:2008zz, Buchalla:2008jp} whereas the present error on $V_{ub}$ is much larger and there is also some tension between the determinations from inclusive and exclusive decay channels. Reduction of this error requires more experimental statistics but---even more so---an improvement of the theoretical prediction of the semileptonic spectra and decay widths.

This is the prime motivation for the study of semileptonic form factors of decays of a heavy meson $H=B,D$ into a light pseudoscalar meson $P=\pi,K$, which are usually defined as
\eq{eq:matrix_element}
\langle P (p) \vert V^\mu \vert H(p_H)\rangle =  \frac{m_H^2 - m_P^2}{q^2} q^\mu
f_0 (q^2) + \left( p^\mu_H + p^\mu - \frac{m_H^2 - m_P^2}{q^2}q^\mu  \right) f_+(q^2)\,.
\en
Here $V^\mu = {\overline q}_2 \gamma^\mu q_1 $ is the vector current in which
$q_1$ ($q_2$) denotes a light (heavy) quark field;
$p$ ($p_H$) is the momentum of the light (heavy) meson with mass $m_P$ ($m_H$), and
$q:=p_H-p$ is the four-momentum transfer.
The $f_0(q^2)$ and $f_+(q^2)$ form factors are dimensionless, real functions of $q^2$ (in the physical region), which encode the strong interaction effects.
They are subject to the kinematic constraint $f_+(0)=f_0(0)$.

In the approximation of massless leptons (which is highly accurate for $\ell=e$ or $\ell=\mu$),
the differential decay rate for the $H\to P\ell\nu_\ell$ process involves $f_+(q^2)$ only:
\eq{eq:differential_decay_rate}
\frac{d \Gamma}{d q^2} =
\frac{G_F^2 \left| V_{q_2 q_1} \right|^{2} }{ 192 \pi^3 m_H^3 }
\left[ \left( m_H^2 + m_P^2 -q^2 \right)^2  - 4 m_H^2 m_P^2 \right]^{3/2} \left| f_+ (q^2) \right|^2\,.
\en

Another motivation for our study is that $f_0(q^2)$ and $f_+(q^2)$ enter as ingredients in the analysis of nonleptonic
two-body decays like $B\to\pi \pi$ and $B\to\pi K$ in the framework of QCD factorization~\cite{Beneke:1999br, Beneke:2001ev},
with the objective to extract CP-violating effects and in particular the angle $\alpha$ of the CKM triangle.
One issue that is especially important in this respect is the question of flavor $\SU(3)$ violation in the form factors of the decay $B \to \pi$ vs. the rare decay $B \to K$.

High-statistics unquenched lattice calculations of $D$-meson (and also
$B$-meson) decay form factors in the kinematic region where the
outgoing light hadron carries little energy (small recoil region) have
been performed recently~\cite{Aubin:2004ej, Okamoto:2005zg, Dalgic:2006dt} and attracted a lot of attention.  Direct simulations
at large recoil, $q^2 \ll m_B^2$, with light hadrons carrying large
momentum of order 2~GeV, prove to be difficult and require a very fine
lattice which is so far not accessible in calculations with dynamical
fermions.  This problem is aggravated by the challenge to consider
heavy quarks which either calls for using effective heavy quark theory
methods or, again, a very fine lattice. In practice, one is forced to
rely on extrapolations from larger momentum transfer $q^2$ and/or
smaller heavy quark masses.  Several extrapolation procedures have
been suggested~\cite{Becirevic:1999kt, Albertus:2005ud, Boyd:1994tt, DescotesGenon:2008hh, Bourrely:2008za} that incorporate constraints from unitarity and the
scaling laws in the heavy quark limit.  Alternatively, $B$-meson form
factors in the region of large recoil have been estimated using
light-cone sum rules~\cite{Balitsky:1989ry, Chernyak:1990ag} (for
recent updates see~\cite{Ball:2004ye, Duplancic:2008ix,
  Duplancic:2008tk, Wu:2009kq}).

In this paper we report on a quenched calculation of semileptonic $H
\to P \ell\nu_\ell$ form factors with lattice spacing $a\sim 0.04$ fm
using nonperturbatively $O(a)$ improved Wilson fermions and $O(a)$
improved currents. On such a fine lattice a relativistic treatment of
the $c$ quark is justified and also the extrapolation to the physical
$b$ quark mass becomes much more reliable compared to similar
calculations on coarser lattices. In addition, we can explore possible
subtleties in approaching the continuum limit in form factor
calculations: In our previous work~\cite{AliKhan:2007tm} we did find
indications for a substantial discretization error in the decay
constants $f_{D_s}$ etc.; similar conclusions have also been reached
in Ref.~\cite{Heitger:2008jq}. This is particularly relevant in view
of the claims of evidence for New Physics from comparison with recent
dynamical simulations---see, e.g. Ref.~\cite{Narison:2008bc} for a
discussion.

On physical grounds, one may expect a nontrivial continuum limit
because form factors at large momentum transfer are determined by the
overlap of very specific kinematic regions in hadron wave functions
(either soft end-point, or small transverse separation).  The common
wisdom that hadron structure is very ``smooth''---and that numerical
simulations on a coarse lattice could thus be sufficient to capture
the continuum physics---may not work in this particular case. This can
be especially important for $\SU(3)$ flavor-violating effects, which
are of major interest for the phenomenology. Inclusion of dynamical
fermions and the approach to the chiral limit are certainly also
relevant problems, but not all issues can be addressed presently
within one calculation.

This work should be viewed as a direct extension of the 
investigation of the APE collaboration in Ref.~\cite{Abada:2000ty},
who performed a quenched calculation with the same nonperturbatively
$O(a)$ improved action and currents. Also their data analysis is
similar. However, they use coarser lattices with $a\sim 0.07$~fm
($\beta=6.2$). On the other hand, the spatial volume of their lattices 
is very close to ours ($L\sim 1.7$~fm). So the main difference lies in
the lattice spacing, and a direct comparison of the results is
possible yielding information on the size of lattice artefacts,
while there is no need for us to perform simulations on a coarser
lattice ourselves.

The presentation is organized as follows. Our strategy is discussed in
detail in Sec.~\ref{sec:simulation_details}. It allows us to run fully
relativistic simulations for values of $m_H$ up to the vicinity of
$m_B$: This is achieved by using a lattice characterized by a very
fine spacing $a$. The extraction of physical quantities from our data
and the final results with the associated error budget are presented
in Sec.~\ref{sec:physical_results}. The final
Sec.~\ref{sec:conclusions} contains a summary and some concluding
remarks. Some technical details and intermediate results of our
calculation are shown in the Appendix. Preliminary results of this
study have been presented in Refs.~\cite{Khan:2007hn, Khan:2009eq}.

\section{Simulation details}
\label{sec:simulation_details}

The lattice study of heavy hadrons is an issue that involves some
delicate technical aspects: The origin of the problem stems from the
fact that, typically, the lattice cutoff is (much) smaller than the
mass of the $B$ meson.

Common strategies to solve this problem are based on heavy quark
effective theory~(HQET), i.e. expanding the relativistic theory in
terms of $m_Q^{-1}$, where $m_Q$ is the mass of the heavy quark. One
can simulate in the static limit~\cite{Eichten:1989zv} or keep
correction terms in the action to simulate at finite
$m_Q$~(non-relativistic QCD or NRQCD)~\cite{Lepage:1992tx}. These
approaches have been employed effectively for studying $B$
physics~(see, for example, Refs.~\cite{Jansen:2008si,
  DellaMorte:2007ij, Hein:2000qu, AliKhan:2001jg}).

However, for smaller quark masses like the $c$ quark in $D$ mesons a
large number of terms in the expansion must be included, making the
simulations less attractive. The Fermilab group developed an approach
which interpolates between the heavy- and light-quark
regimes~\cite{ElKhadra:1996mp}. The coefficients accompanying
each term in the action are functions of the quark mass and in
practice, normally, the lowest-level action is used. This corresponds
to using the $O(a)$ improved relativistic action with a
re-interpretation of the results. Except for
HQET~\cite{Heitger:2003xg, DellaMorte:2006sv}, the associated
renormalization constants for these approaches are only known
perturbatively.

We reduce the uncertainties related to the extrapolation to the
physical heavy meson mass by using lattices with a small spacing in
conjunction with a non-perturbatively improved $O(a)$ relativistic
quark action. This theoretically clean approach enables one to get
sufficiently close to the mass of the physical $B$ meson, so that the
heavy-quark extrapolation is short-ranged. In addition, in the region
of the $D$ meson mass, the discretization errors are reduced to around
$1\%$, see Section~\ref{sys}.

\begin{table}[h]
\centering
\phantom{-------}
\begin{tabular}{|cc|}  \hline
$L^3 \times T$ & $40^3 \times 80$ \\
$\beta$ & $6.6$ \\
lattice spacing $a$ & $0.04$~fm \\
physical hypervolume & $1.6^3 \times 3.2 $~fm$^4$ \\
$a^{-1}$ & $4.97$ GeV \\
\# of configurations & $114$ \\
\hline
$\kappa_{critical}$ & $ 0.135472(11) $ \\
$\kappa_{heavy}$ & $0.13$, $0.129$, $0.121$, $0.115$ \\
$\kappa_{light}$ & $0.13519$, $0.13498$, $0.13472$ \\
$m_P$ & $526$~MeV, $690$~MeV, $856$~MeV \\
$c_{SW}$ & $1.467$ \\
$Z_V$ & $0.8118$
 \\
$b_V$ & $1.356$
\\
$c_V$ & $-0.0874$
\\
 \hline
\end{tabular}
\phantom{-------}
\caption{Parameters of the lattice calculation (see the text for the definition of the various quantities).}\label{tab:simulation_information}
\end{table}

Table~\ref{tab:simulation_information} summarizes basic technical
information about our study. We use the standard Wilson gauge action
to generate quenched configurations with the coupling parameter
$\beta=6.6$. For this parameter choice, the lattice spacing in
physical units determined from Ref.~\cite{Necco:2001xg} using Sommer's
parameter $r_0=0.5$~fm is $a=0.04$~fm. Our calculation is based on the
$O(a)$ improved clover formulation for the quark
fields~\cite{Sheikholeslami:1985ij}, with the nonperturbative value of
the clover coefficient $c_{SW}$ taken from Ref.~\cite{Luscher:1996ug}.
We use $O(a)$ improved definitions of the vector currents in the
form~\cite{Luscher:1996jn} \eq{eq:oa_improved_vector_current} V_\mu =
Z_V \left[ 1+ b_V \frac{a m_{q_2}+ a m_{q_1}}{2} \right] \left(
  {\overline q}_2 \gamma_\mu q_1 + i a c_V \partial_\nu {\overline
    q}_2 \sigma_{\mu\nu} q_1 \right) \en with $\sigma_{\mu
  \nu}=\frac{i}{2} \left[ \gamma_\mu,\gamma_\nu \right]$. The
renormalization factor $Z_V$, the improvement coefficient $b_V$ as
well as $c_V$ are known nonperturbatively~\cite{Bakeyev:2003ff,
  Guagnelli:1997db, Pleiter_thesis}. All statistical errors are evaluated through a
bootstrap procedure with $500$ bootstrap samples. We consider three
hopping parameters corresponding to ``light'' quarks, $\kappa_{light}$ (the corresponding masses of the light pseudoscalar meson states $m_P$ are also given in Table~\ref{tab:simulation_information}), and four hopping
parameters, $\kappa_{heavy}$, corresponding to ``heavy'' quarks; in particular,
$\kappa=0.13498$ and $\kappa=0.129$ are found to correspond to quark
masses close to the physical strange and charm quark mass, respectively.

\begin{figure}
\centerline{\includegraphics[width=.50\textwidth]{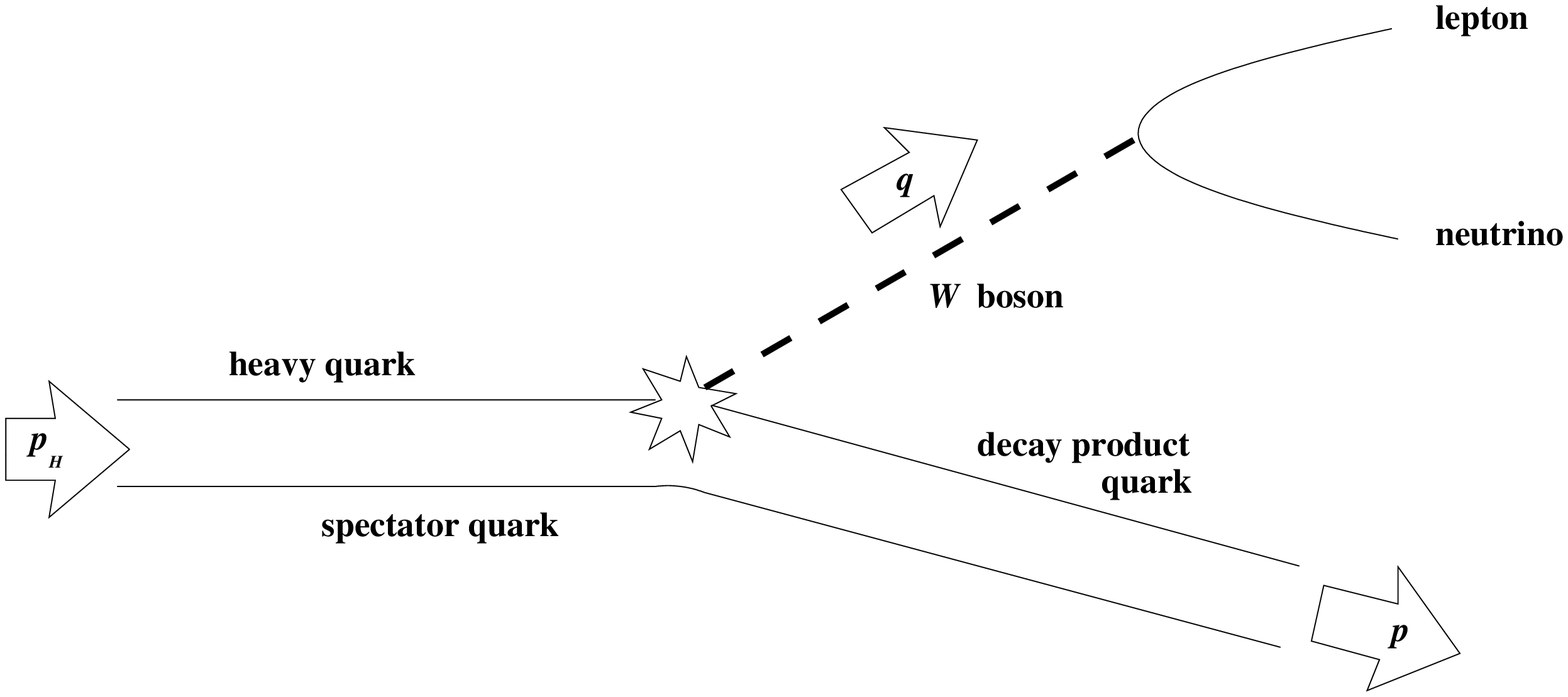}\hspace{1cm}\includegraphics[width=.35\textwidth]{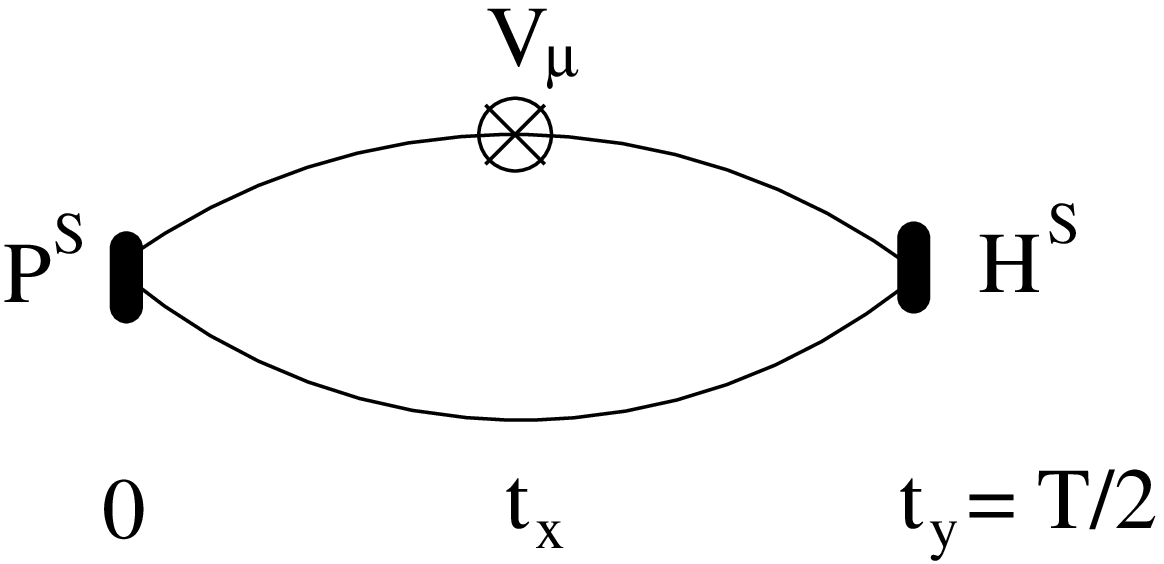}}
\caption{Diagram representing the semileptonic decay of a heavy-light pseudoscalar meson to a light pseudoscalar meson (left panel). A schematic representation of the corresponding three-point correlator calculated on the lattice is also shown (right panel).}
  \label{fig:diagram}
\end{figure}

The extraction of the matrix element appearing in Eq.~(\ref{eq:matrix_element}) from the lattice can be done by considering the large time behavior of three-point correlation functions $C_\mu^{(3)}(0,t_x,t_y)$ for a pseudoscalar light meson sink at time $t=0$, a vector current at time $t_x$, and a pseudoscalar heavy-light meson source at time $t_y = T/2$ (see Fig.~\ref{fig:diagram}):
\eq{eq:three_point_function}
C_\mu^{(3)}(0,t_x,t_y) = \sum_{\vec{x},\vec{y}}
e^{-i\vec{p}_H\cdot \vec{y}}e^{i\vec{q}\cdot \vec{x}}
\langle H^S(\vec{y},t_y)V_\mu(\vec{x},t_x)P^S(0) \rangle \:.
\en
Here, $H^S$ and $P^S$ are Jacobi-smeared operators of the form $\overline{q}_{h} \gamma_5q_{s}$ and $\overline{q}_{l}\gamma_5q_{s}$, respectively; $q_h$ denotes the heavy quark, $q_l$ is the decay-product quark, while $q_{s}$ is the ``spectator'' quark.

For sufficiently large time separations (i.e. $0 \ll t_x \ll T/2 $ or $T/2 \ll t_x \ll T$), $C_\mu^{(3)}(0,t_x,t_y)$ behaves as:
\eq{eq:large_time_behavior_of_C_3}
C_\mu^{(3)}(0, t_x,t_y)  \longrightarrow
\left\{
\begin{array}{lr}
\frac{Z^S_H Z^S}{4E_H E}e^{-E t_x}
e^{-E_H (t_y-t_x)}\;\langle H (p_H) |V_\mu| P(p) \rangle & \mbox{for } \;
t_x < T/2 \\
{\phantom{\huge{l}}} & \\
\pm \frac{Z^S_H Z^S }{4 E_H E}
e^{-E (T-t_x)}e^{-E_H (t_x-t_y)}\;
\langle H(p_H)|V_\mu| P(p) \rangle & \mbox{for } \;  t_x > T/2
\end{array}
\right. \:, \en with $Z^S_H = |\langle 0|H^S|H(p_H)\rangle|$ and $Z^S
= |\langle 0|P^S|P(p)\rangle|$, while $E$ ($E_H$) denotes the energy
of the light (heavy) meson. To extract the matrix elements we divide
the three-point functions by the prefactors, which are extracted from
fits to smeared-smeared two-point functions. The matrix element is
then obtained by fitting this result to a constant, in an appropriate
time range where a clear plateau forms (for example, for $12 \le t_x \le
28$).

We consider three-point functions associated with different
combinations of the momenta $p$ and $p_H$, which are listed in
Table~\ref{tab:momentum_combinations}. In particular, we focus our
attention onto three-momenta of modulus $0$ and $1$ [in units of
$2\pi/(aL)$], since they yield the most precise signal, restricting
ourselves to the cases where $\vec{p}$ and $\vec{p}_H$ lie in the same
direction. Thus we measure directly $5$ different values for the form
factors, for every $\kappa_{light}$ and $\kappa_{heavy}$ combination.
  The full form factors can then be constructed from the data points
  obtained this way, by making an ansatz for the functional form of
  $f_0(q^2)$ and $f_+(q^2)$.

\begin{table}[h]
\centering
\phantom{---------------}
\begin{tabular}{|ccc|}  \hline
$\vec{p}_H$ & $\vec{p}$ & $\vec{q} $ \\
\hline
 $(0,0,0)$ &  $(1,0,0)$ & $(-1,0,0)$  \\
 $(1,0,0)$ &  $(-1,0,0)$ & $(2,0,0)$  \\
 $(0,0,0)$ &  $(0,0,0)$ & $(0,0,0)$  \\
 $(1,0,0)$ &  $(0,0,0)$ & $(1,0,0)$  \\
 $(1,0,0)$ &  $(1,0,0)$ & $(0,0,0)$  \\
 \hline
\end{tabular}
\phantom{---------------}
\caption{Momentum combinations considered in the analysis of the three-point functions, in units of $2\pi/(aL)$.}\label{tab:momentum_combinations}
\end{table}

In the present work, we fit our data with the parametrization proposed
by Be\'cirevi\'c and Kaidalov~\cite{Becirevic:1999kt}:
\eq{eq:Becirevic_Kaidalov_parametrization} f_0( q^2 ) = \frac{c_{BK}
  \cdot (1-\alpha)}{1- \tilde q^2/\beta} \:, \qquad f_+( q^2 ) =
\frac{c_{BK} \cdot (1-\alpha)}{(1- \tilde q^2) (1- \alpha \tilde q^2
  )}, \en where $\tilde q :=q/m_{H^\star}$, $m_{H^\star}$ being the
mass of the lightest heavy-light vector meson.

The parametrization for the form factors given in
Eq.~(\ref{eq:Becirevic_Kaidalov_parametrization}) accounts for the
basic properties that come from the heavy-quark scaling laws in the
limits of large and small recoil and also satisfies the
proportionality relation derived in Ref.~\cite{Charles:1998dr}. It is
also consistent with the trivial requirement that the l.h.s.\ of
Eq.~(\ref{eq:matrix_element}) be finite for vanishing momentum transfer,
which implies $f_0(0)=f_+(0)$.  The results that
we obtained for the three parameters entering
Eq.~(\ref{eq:Becirevic_Kaidalov_parametrization}) from a simultaneous
fit to $f_0$ and $f_+$ are presented in the Appendix.

Some alternative ans\"atze for the functional form of $f_+(q^2)$ were
proposed in Refs.~\cite{Ball:2004ye, Albertus:2005ud, Boyd:1994tt} and
are discussed in Ref.~\cite{Ball:2006jz}: They yield results
essentially compatible with each other and with the
Be\'cirevi\'c-Kaidalov parametrization
Eq.~(\ref{eq:Becirevic_Kaidalov_parametrization}).  More recently,
Bourrely, Caprini and Lellouch~\cite{Bourrely:2008za} discussed the
representation of $f_+(q^2)$ as a (truncated) power series in terms of an auxiliary variable $z$. A similar parametrization
has also been recently used by the Fermilab Lattice and MILC
collaborations, see Refs.~\cite{Vandewater_lattice_2008, Bailey:2008wp} for a discussion.

\section{Extraction of physical results}
\label{sec:physical_results}

\begin{figure}
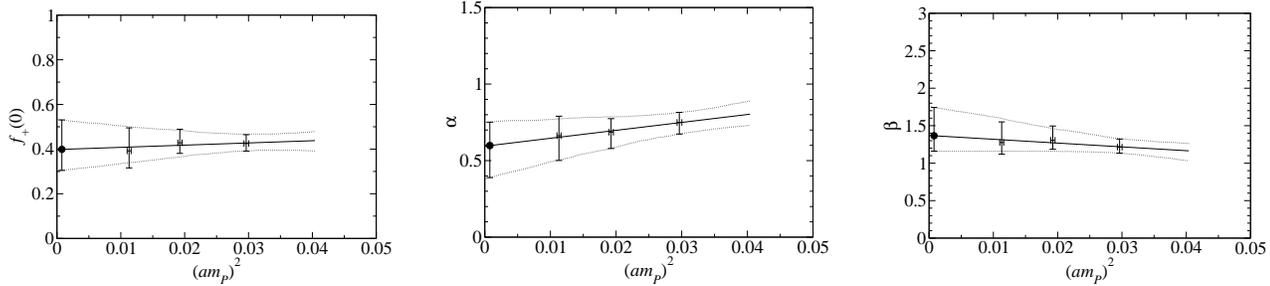

\centerline{
\includegraphics[width=.30\textwidth]{chirally_extrapolated_fplusatzero_parameter_plot_from_kdec_LIGHT_kh_11500_kspec_LIGHT.eps}\hfill \includegraphics[width=.30\textwidth]{chirally_extrapolated_alpha_parameter_plot_from_kdec_LIGHT_kh_11500_kspec_LIGHT.eps}\hfill
\includegraphics[width=.30\textwidth]{chirally_extrapolated_beta_parameter_plot_from_kdec_LIGHT_kh_11500_kspec_LIGHT.eps}
}
\caption{Extrapolation of the Be\'cirevi\'c--Kaidalov parameters to
  the chiral limit, for decays to a pion, at a fixed value
  $\kappa_{heavy}=0.115$. The parameters obtained for $\kappa_{decay
    \; product}=\kappa_{spectator}$ are extrapolated linearly in 
  $m_P^2$. The extrapolated values are shown as the full black
  dots.}
\label{fig:pion_chiral_extrapolation_kheavy_11500}
\end{figure}

In order to extract physical results from our simulations, we follow a
method analogous to Ref.~\cite{Abada:2000ty}. We first perform a
chiral extrapolation in the light quark masses. For a given quantity
$\Phi$ [one of the BK parameters appearing in
  Eq.~(\ref{eq:Becirevic_Kaidalov_parametrization})], the
extrapolation relevant for decays to a pion is performed as follows:
We fit the results obtained at different values of the mass of the
pseudoscalar state linearly in $m_P^2$,
\eq{eq:light_quark_extrapolation} \Phi = c_0 + c_1 \cdot m_P^2 \;, \en
and extrapolate to $m_P^2=m_\pi^2$, where $m_\pi$ is the mass of the
physical pion. Examples of the
extrapolations are shown in
Figure~\ref{fig:pion_chiral_extrapolation_kheavy_11500} for the case
of $\Phi=f_+(0)$, $\alpha$ and $\beta$ at $\kappa_{heavy}=0.115$. On the other hand, for decays to a kaon, we hold the
hopping parameter of one of the two final quarks fixed to
$\kappa=0.13498$, which, for our configurations, corresponds to the
physical strange quark at a high level of precision~\cite{AliKhan:2007tm}, 
and perform a short-ranged extrapolation of the curve
obtained from the linear fit in $m_P^2$ to the square of the mass of
the physical $K$ meson.

Then we perform the interpolation to the physical $c$ quark mass in
terms of a heavy-quark expansion for the $D$ (or $D_s$) meson decays,
or the extrapolation to the physical $b$ quark mass for the $B$ (or
$B_s$) meson.  For our data, the extrapolation of the heavy quark mass
to the physical $b$ mass is short-ranged: for the heaviest
$\kappa_{heavy}=0.115$, it turns out that the inverse of the
pseudoscalar meson mass (with the light quark mass already chirally
extrapolated to its physical value) is about $m_H^{-1} = 0.243$
GeV$^{-1}$, to be compared with $m_B^{-1} =0.189$ GeV$^{-1}$ for the
physical $B$ meson.  The extrapolation can be performed by taking
advantage of the fact that, in the infinitely heavy quark limit, the
Be\'cirevi\'c--Kaidalov parameters appearing in
Eq.~(\ref{eq:Becirevic_Kaidalov_parametrization}) enjoy certain
scaling relations: $c_{BK} \sqrt{m_H}$, $(\beta-1) m_H$ and
$(1-\alpha) m_H$ are expected to become constant in the
$m_H \rightarrow \infty$ limit. For finite $m_H$, one can parametrize
the scaling deviations in powers of $m_H^{-1}$:
\eq{eq:linear_heavy_quark_expansion} \varphi = l_0 + l_1\cdot m_H^{-1}
+ l_2\cdot m_H^{-2} + \ldots \en where $\varphi \in \{ c_{BK}
\sqrt{m_H}, \; (\beta-1) \cdot m_H, \; (1-\alpha) \cdot m_H \}$. Note
that, since $f_+(0)=c_{BK} \cdot (1-\alpha)$, one can also use
$\varphi = f_+(0) \cdot m_H^{3/2}$---which was, in fact, our choice.

\begin{figure}
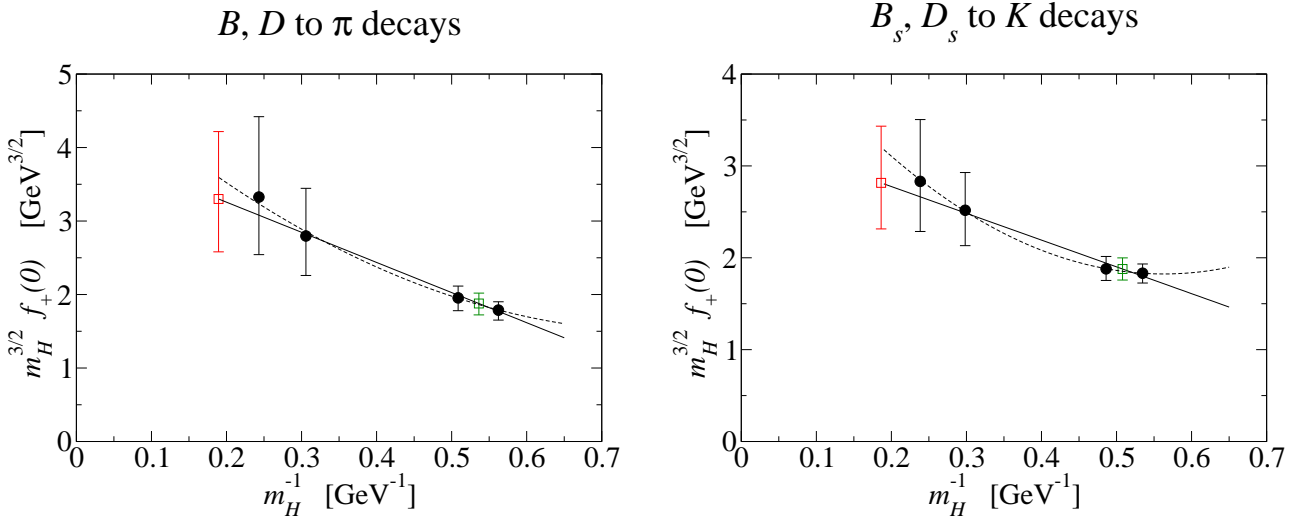

\centerline{\includegraphics[width=.48\textwidth]{decay_from_H_to_pi.eps}\hspace{.04\textwidth}\includegraphics[width=.48\textwidth]{decay_from_H_s_to_K.eps}}
\caption{Left panel: The green (red) squares denote the interpolated
  (extrapolated) form factor $m_H^{3/2}f_+(0)$ to the physical $D$ ($B$) meson,
  for a decay to a pion, using a linear fit in $1/m_H$ (solid line).
  A quadratic fit to the data is also shown (dashed line). Right
  panel: the results for the case of decay of a $D_s$ ($B_s$)
  meson into a kaon.}
  \label{fig:decays}
\end{figure}

The extrapolation of $m_H^{3/2} f_+(0)$ is presented in
Figure~\ref{fig:decays}. The figure clearly shows the advantage of
simulating on a fine lattice, which allows us to probe a mass range
very close to the physical $B$ meson mass. We compare the results
obtained from an extrapolation to the inverse of the physical $B$
meson mass using either a first- or a second-order polynomial in
$m_H^{-1}$ for the fit function, finding consistency (within
error bars), for all decays. The corresponding fit results are listed
in Table~\ref{tab:linear_vs_quadratic_fit_results}. In the following
we refer to this first method as the ``coefficient extrapolation''
method.

\begin{table}
\centering
\begin{tabular}{|c|c|c|c|c|c|}\hline
        Decay          & fit       & $l_0$                    & $  l_1 $                 & $ l_2$
   & $ \chi^2/$d.o.f.  \\ \hline
$B,D\rightarrow \pi$   & linear    & $ 4.1^{+1.3}_{-1.0}    $ & $  -4.1^{+1.6}_{-2.3} $  & --
   & $ 0.1377/2     $    \\
                       & quadratic & $ 5.1^{+2.9}_{-2.1}    $ & $  -9.3^{+6.9}_{-9.6} $  & $ 5.9^{+8.5}_{-5.9}
$  & $ 0.021/1      $    \\\hline
$B,D\rightarrow K  $   & linear    & $ 4.9^{+1.1}_{-0.9}    $ & $  -5.4^{+1.5}_{-1.9} $  & --
   & $ 0.3247/2     $    \\
                       & quadratic & $ 6.3^{+2.4}_{-1.9}    $ & $ -12.2^{+6.3}_{-8.2} $  & $ 7.7^{+7.3}_{-5.4}
$  & $ 0.03813/1    $    \\\hline
$B_s,D_s\rightarrow K$ & linear    & $ 3.4^{+0.9}_{-0.8} $ & $  -2.9^{+1.3}_{-1.7} $  & --
& $ 0.4025/2     $    \\
                       & quadratic & $ 4.9^{+1.8}_{-1.3}    $ & $ -11.0^{+4.6}_{-6.0} $  & $ 9.7^{+5.4}_{-4.6}
$  & $ 0.001888/1   $   \\\hline
\end{tabular}
\caption{ The coefficients obtained from the fits to $m_H^{3/2}
  f_+(0)$ in powers of $m_H^{-1}$ according to
  Eq.~(\protect\ref{eq:linear_heavy_quark_expansion}) for different
  decays.
\label{tab:linear_vs_quadratic_fit_results}}
\end{table}

An alternative method to extract the physical form factors from the
lattice data was proposed by the UKQCD
collaboration~\cite{Bowler:1999xn}. It consists of performing the
chiral and heavy quark extrapolations at fixed $v\cdot p = ( m_H^2 +
m_P^2 - q^2 )/(2 m_H)$, where $v$ is the four-velocity
  of the heavy meson and $p$ is the four-momentum of the light
  meson. The following steps are performed:
\begin{enumerate}
\item fit of the form factors measured from the lattice simulations to
  the parametrization in
  Eq.~(\ref{eq:Becirevic_Kaidalov_parametrization});
\item interpolation of the form factors at given values
  of $v \cdot p$ within the range of
  simulated data;
\item chiral extrapolation of the points thus obtained, via a linear
  extrapolation in $m_P^2$ to either $m_\pi^2$ or $m_K^2$ (as
  described above);
\item linear or quadratic extrapolation in $m_H^{-1}$ to the inverse
  of the physical heavy meson mass for the quantities:
  \eq{eq:scaling_quantities_heavy_extrap_UKQCD_method} \left[
    \frac{\alpha_s(m_B)}{\alpha_s(m_H)} \right]^{-\frac{\tilde
      \gamma_0}{2 \beta_0}} f_0(v \cdot p) \sqrt{m_H} \,,\qquad \qquad
  \left[ \frac{\alpha_s(m_B)}{\alpha_s(m_H)} \right]^{-\frac{\tilde
      \gamma_0}{2 \beta_0}} \frac{f_+(v \cdot p)}{\sqrt{m_H}} \en
  which enjoy scaling relations at fixed $v \cdot
  p$~\cite{Isgur:1990kf, Neubert:1993za}. Here, $\beta_0$ is the first
  $\beta$-function coefficient, while $\tilde \gamma_0=-4$ denotes the
  leading-order coefficient of the anomalous dimension for the vector
  current in HQET. It yields a (subleading) logarithmic dependence on
  $m_H$---see also Refs.~\cite{Abada:2000ty, Bowler:1999xn} for
  further details;
\item final fit of the points thus obtained to the parameterization in Eq.~(\ref{eq:Becirevic_Kaidalov_parametrization}).
\end{enumerate}
For comparison, we also calculate the physical form factors using
this alternative approach, finding consistent results. This is
illustrated in Table~\ref{tab:final_results} which summarizes the
results for $f_+(0)$ from both methods.

\begin{figure}
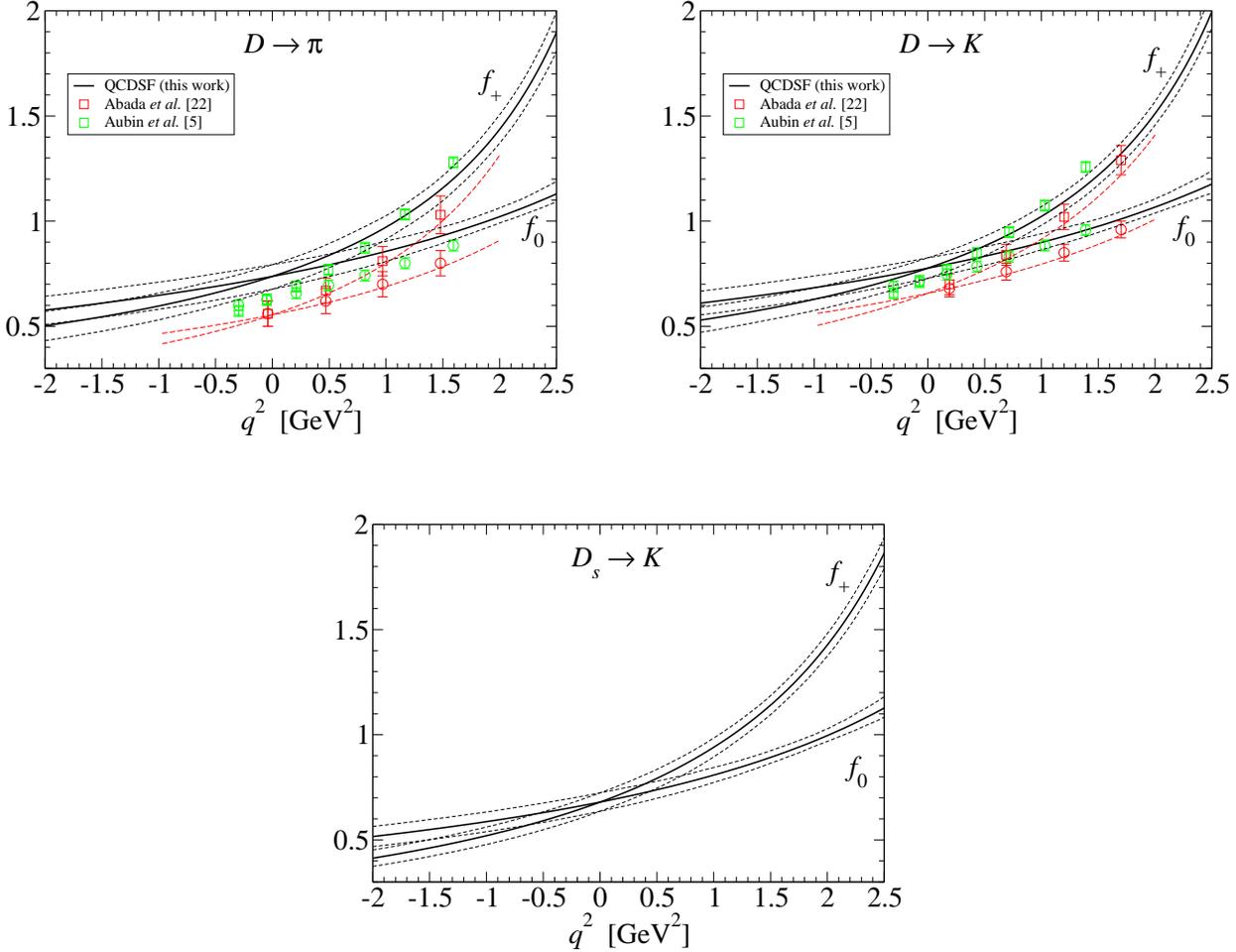

\centerline{\includegraphics[width=.45\textwidth]{compare_D_to_pi_extrapolated_curves.eps}\hspace{1cm}
\includegraphics[width=.45\textwidth]{compare_D_to_K_extrapolated_curves.eps}}
\vspace{10mm}
\centerline{\includegraphics[width=.45\textwidth]{D_s_to_K_extrapolated_curves.eps}}
\caption{Physical form factors for $D$ and $D_s$ decays as a function
  of $q^2$ from this work and other quenched and dynamical studies.
  The solid black lines are the form factors obtained from the
  coefficient extrapolation method where
  Eq.~(\protect\ref{eq:linear_heavy_quark_expansion}) has been
  truncated at $O(m_H^{-1})$, while the dashed black lines indicate the
  error on the form factors. The range of $v\cdot p$ values achieved in our simulations approximately corresponds to $-1.5$~GeV$^2 \lesssim q^2 \lesssim$~$2$~GeV$^2$. The dashed red lines are the results for the coefficient extrapolation method from Ref.~\protect\cite{Abada:2000ty}. The open red squares and circles  are their results obtained using the UKQCD method.}
  \label{fig:charm_decays_extrapolated_curves}
\end{figure}

\begin{figure}
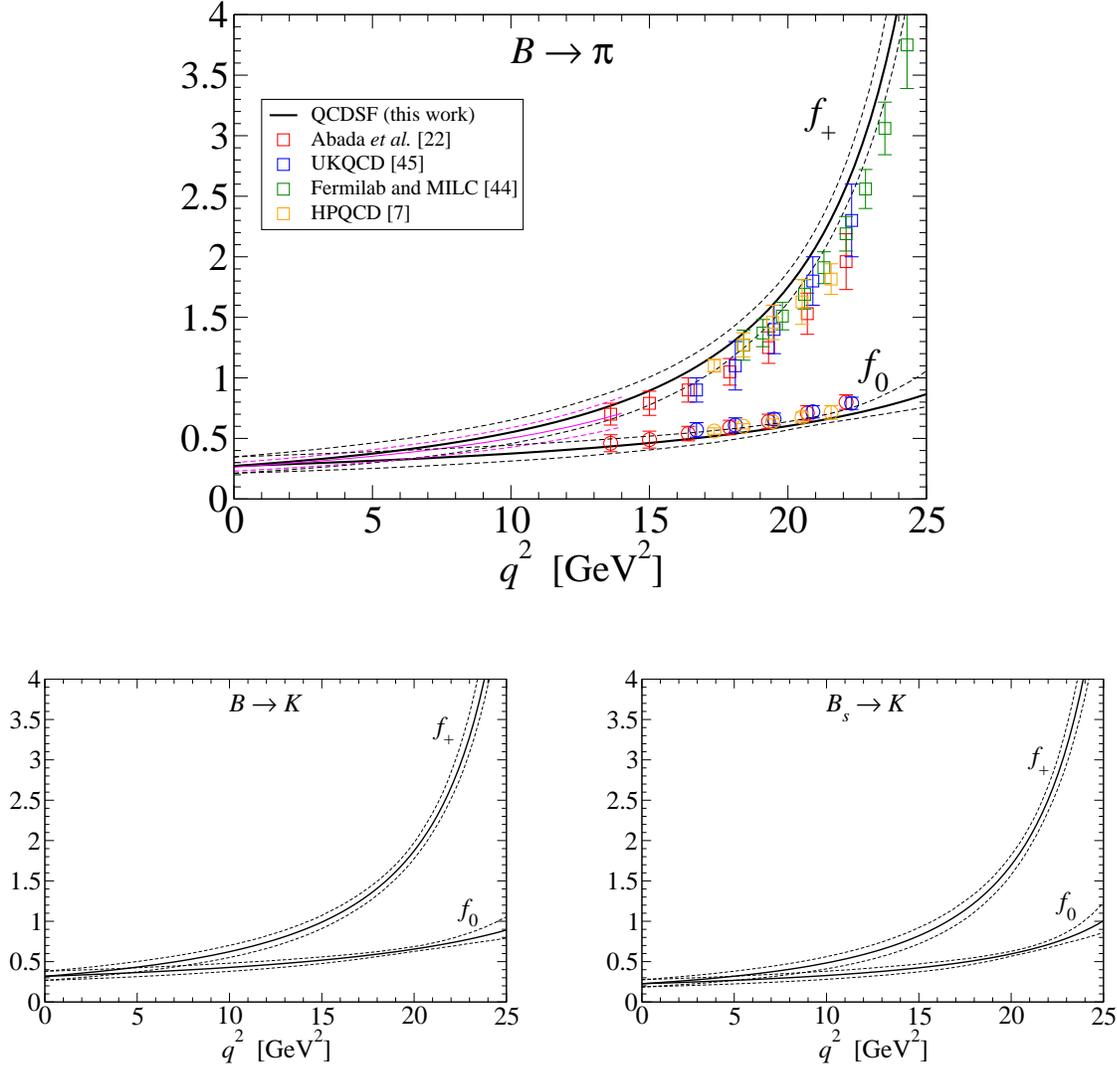

\centerline{\includegraphics[width=.60\textwidth]{compare_B_to_pi_extrapolated_curves.eps}}
\vspace{10mm}
\centerline{\includegraphics[width=.40\textwidth]{B_to_K_extrapolated_curves.eps}\hspace{1cm}
\includegraphics[width=.40\textwidth]{B_s_to_K_extrapolated_curves.eps}}
\caption{Same as in
  Fig.~\protect\ref{fig:charm_decays_extrapolated_curves}, but for $B$
  and $B_s$ decays; in this case, the $v\cdot p$ values of our
  simulations are in the range $14$~GeV$^2 \lesssim q^2 \lesssim$~$23$~GeV$^2$. For $B\rightarrow \pi$, the dashed and solid
  magenta lines in the range $q^2=0-14$~GeV$^2$ indicate the
  prediction from light-cone sum
  rules~\protect\cite{Ball:2004ye,Duplancic:2008ix}.}
  \label{fig:beauty_decays_extrapolated_curves}
\end{figure}

For our final results we take those obtained from the coefficient
extrapolation method. We found this method to be superior in our case
as the UKQCD method suffered from the fact that there was only a small
region of overlap in the ranges of $v\cdot p$ for the form factors at
different $\kappa_{light}$ and $\kappa_{heavy}$.
In addition, since the data can be fitted with both a linear and quadratic function in $m_H^{-1}$, we use the linear fits for the central values and statistical errors and use the differences in the results from the linear and quadratic fit to estimate the systematic errors, as discussed in the next section. Our
results for the form factors at finite $q^2$ are shown in Figs.~\ref{fig:charm_decays_extrapolated_curves} and~\ref{fig:beauty_decays_extrapolated_curves}.

\begin{table}[h]
\centering
\begin{tabular}{|c|c|c|c|c|} \hline
 & \multicolumn{2}{|c|}{Coefficient extrapolation} & \multicolumn{2}{|c|}{UKQCD method} \\
\hline
Decay & linear in $m_H^{-1}$ & quadratic in $m_H^{-1}$ & linear in $m_H^{-1}$ & quadratic in $m_H^{-1}$ \\
\hline
 $ D \rightarrow \pi $  & $ 0.74^{+6}_{-6} $ & $ 0.73^{+5}_{-6} $ & $ 0.69^{+5}_{-5} $ & $ 0.69^{+5}_{-6} $ \\
 $ D \rightarrow   K $  & $ 0.78^{+5}_{-5} $ & $ 0.77^{+5}_{-5} $ & $ 0.75^{+4}_{-5} $ & $ 0.75^{+4}_{-5} $ \\
 $ D_s \rightarrow K $  & $ 0.68^{+4}_{-4} $ & $ 0.67^{+4}_{-4} $ & $ 0.68^{+4}_{-4} $ & $ 0.67^{+4}_{-4} $ \\
 $ B \rightarrow \pi $  & $ 0.27^{+8}_{-6} $ & $0.30^{+11}_{-8} $ & $ 0.29^{+13}_{-8}$ & $ 0.31^{+15}_{-10}$ \\
 $ B \rightarrow   K $  & $ 0.32^{+6}_{-5} $ & $ 0.35^{+9}_{-8} $ & $ 0.35^{+11}_{-8}$ & $ 0.34^{+12}_{-9} $ \\
 $ B_s \rightarrow K $  & $ 0.23^{+5}_{-4} $ & $ 0.26^{+7}_{-5} $ & $ 0.23^{+6}_{-5} $ & $ 0.27^{+8}_{-6} $ \\
 \hline
\end{tabular}
\caption{ Final results for the physical values of the $f_+(0)$ form
  factor, for different decays, with statistical errors only. We compare the results obtained from the coefficient extrapolation and UKQCD methods as well as different truncations of the heavy quark expansion when extrapolating or interpolating in $m_H^{-1}$.} \label{tab:final_results}
\end{table}

\subsection{Systematic uncertainties}
\label{sys}
Systematic uncertainties affecting our lattice calculation include: the quenched approximation, the method to set the quark masses, the chiral extrapolation for the light
quarks, discretization effects, the extrapolation (interpolation) of the heavy quark to the
physical $b$ ($c$) mass, finite volume effects, uncertainties in the renormalization coefficients, and effects related to the model dependence for $f_{0,+}(q^2)$.
Let us now consider each source of error in turn.

{\bf Quenched approximation:} the size of the error this approximation
introduces is not known. However, one can take as an estimate the
variation in the results if different quantities are used to set the
scale. In the quenched approximation different determinations of the
lattice spacing vary by approximately $10\%$~\cite{Aoki:1999yr}. By repeating the full analysis, we find that varying the lattice spacing by $10\%$ induces an uncertainty of approximately $2\%$ for the $D \rightarrow \pi$ decay, and of approximately $12\%$ for $B \rightarrow \pi$.

{\bf Setting the quark masses:} we use the $\kappa$ values corresponding to the light~($u/d$) and strange quarks determined in Ref.~\cite{AliKhan:2007tm}: $\kappa_l=0.135456(10)$ and $\kappa_s=0.134981(9)$. The uncertainty in these determinations leads to a very small uncertainty in the form factors. For the $c$ and $b$ quarks we do not quote the corresponding $\kappa$ values. We interpolate (or extrapolate) our results directly to the physical masses of the pseudoscalar heavy-light states. The resulting uncertainty is determined by the statistical errors of the masses used for the interpolation (or extrapolation). The latter are found to contribute only a negligible amount to the overall systematic uncertainty.

{\bf Chiral extrapolation:} the method we used to perform the chiral
extrapolation of our simulation results is discussed above. Note that
the use of a large lattice practically constrains us to use only a few
and relatively large values for the light quark mass (so that the
masses of our lightest pseudoscalar mesons are far from the physical
pion mass).  However, as the examples in
Figure~\ref{fig:pion_chiral_extrapolation_kheavy_11500} show, the
dependence of our results on the light quark mass is rather mild. So
the size of the uncertainty arising from the chiral extrapolation
though difficult to estimate is unlikely to be large.

{\bf Discretization effects:} as it was already remarked above, the
leading discretization effects in our calculation are reduced to
$O(a^2)$; given that our lattice is very fine ($a=0.04$~fm), the
associated systematic error can be estimated to be of the order of
$1\%$ ($10\%$) for the decays of charmed (beautiful) mesons~\cite{AliKhan:2007tm}.

{\bf Extrapolation/interpolation of the heavy quark:} our data can be fitted to both a linear and quadratic function in $m_H^{-1}$ with a reasonable $\chi^2$. We use the results for the linear fit for our final results and the difference between the linear and quadratic fit as an indication of the systematic uncertainty. This leads to approximately a $1\%$ uncertainty for $D$ decays and $8\%$ uncertainty for $B$ decays.

{\bf Finite volume effects:} for our calculation, finite-volume
effects are not expected to be severe; in particular, the correlation
length associated with the lightest pseudoscalar state that we
simulated (for $\kappa_{light}=0.13519$) corresponds to approximately
$9$ lattice spacings, which is more than four times shorter than the
spatial extent of our lattice. Systematic infrared effects can thus be
quantified around $1$--$2\%$. This is comparable with the estimate of Ref.~\cite{Bailey:2008wp}, in which, using chiral perturbation theory~\cite{Aubin:2007mc, Arndt:2004bg}, the finite volume effects for their calculation with $2+1$ flavors of staggered quarks and values of $m_P L$ between $4$ and $6$ are estimated to be less than $1\%$.

{\bf Renormalization coefficients:} the uncertainty associated with the $Z_V$ coefficient, as determined in Ref.~\cite{Bakeyev:2003ff} for the quenched case, is about $0.5\%$. The same article also quotes a $1\%$ uncertainty for $b_V$, which induces an error about $1\%$ for decays of $D$ mesons and about $3\%$ for $B$ mesons. Concerning $c_V$, a look at the results displayed in Fig.~2 of Ref.~\cite{Guagnelli:1997db} would suggest that the relative error in the region of interest ($g_0^2 \simeq 0.91$) may be quite large, around $30\%$; however, it should be noted that $c_V$ itself is a relatively small number, of the order of $9\%$, and the impact of the uncertainty on $c_V$ on our results is about $1\%$ ($2\%$) for decays of $D$ (respectively: $B$) mesons.

{\bf Model dependence:} finally, the systematic effect related to the ansatz to parametrize the form factors was estimated in
Ref.~\cite{Ball:2006jz}, through a comparison of different functional forms that satisfy analogous physical requirements. For the $B \rightarrow \pi$ decay, it turns out to be of the order of $2\%$.

Combining the systematic errors in quadrature we arrive at an overall error of $5\%$ for $D$ decays and about $18\%$ for $B$ decays.

\subsection{Comparison with previous results}

Our results can be compared to other lattice calculations of these
quantities and also with results of light-cone sum rules
(LCSR)~\cite{Balitsky:1989ry,
  Chernyak:1990ag}. Table~\ref{tab:comparison_with_other_works}
summarizes the comparison for $f_+(0)$, while for finite $q^2$ the
form factors from other studies are displayed in
Figs.~\ref{fig:charm_decays_extrapolated_curves}
and~\ref{fig:beauty_decays_extrapolated_curves}. In the following we
discuss in detail the comparison with these works, highlighting the
advantages and limitations of the different approaches, as well as the
possible sources of discrepancies.

Our results can be closely compared with those obtained by the APE
collaboration in Ref.~\cite{Abada:2000ty}, reporting a calculation
very similar to ours. They worked in the quenched approximation, using
the same non-pertubatively $O(a)$ improved action and currents and a
similar analysis; on the other hand, their simulations were performed
on a coarser lattice, with $\beta=6.2$, yielding a lattice spacing
$a=0.07$~fm, or $a^{-1} \simeq 2.7$~GeV. The table and figures show
that their values for the form factors lie around
$3\sigma$~($D\rightarrow \pi$) and $2.5\sigma$~($D\rightarrow K$)
below our results, in terms of the statistical errors, in the region
of $q^2=0$.  If we adjust the APE results to be consistent with
setting the lattice spacing using $r_0$ instead of the mass of the
$K^*$~(used in Ref.~\cite{Abada:2000ty}), the discrepancy reduces
slightly, down to roughly $2.5\sigma$~($D\rightarrow \pi$) and
$2\sigma$~($D\rightarrow K$).  Assuming that $O(a^2)$ errors are the
dominant source of the discrepancy, the difference in the results of
the two studies is consistent with an upper limit on the
discretization errors of approximately $0.08$, or slightly above
$1\sigma$ in our results for $f_+^{D\rightarrow \pi}(0)$ and $0.23$ or
$3-4\sigma$ in the APE results.

For $B$ decays we are not able to make such a close comparison,
because the study in Ref.~\cite{Abada:2000ty} extrapolates to the $B$
meson from results in the region of $1.7-2.6$~GeV for the heavy-light
pseudoscalar meson mass. Although one would expect larger
discretization effects for the $B$ decay form factors, we find close
agreement between our values and those from the APE
collaboration. However, we should point out that any potential
discrepancy may be masked by the long-ranged extrapolation in the
heavy quark mass.

Several unquenched calculations have been performed recently, which
are based on the MILC $N_f=2+1$ dynamical rooted staggered fermions
configurations~\cite{Bernard:2001av}. Results are available from joint
works from the Fermilab, MILC and HPQCD collaborations for $D$
decays~\cite{Aubin:2004ej}, and from Fermilab and
MILC~\cite{Okamoto:2004xg, Bailey:2008wp} and~(separately)
HPQCD~\cite{Dalgic:2006dt} for $B$ decays.
% are shown in
%Table~\ref{tab:comparison_with_other_works} and
%Figs.~\ref{fig:charm_decays_extrapolated_curves}
%and~\ref{fig:beauty_decays_extrapolated_curves}. 
These results were
obtained using the MILC ``coarse'' lattices with $a=0.12$~fm for $D$
decays and including a finer lattice with $a=0.09$~fm for the $B$
decays. While these lattices are much coarser than those used in both
our and the APE study a detailed analysis of the chiral
extrapolation was possible through the use of 5 light quark masses for
the $0.12$~fm lattice~(only two values were used for $a=0.09$~fm).

The Fermilab, MILC and HPQCD joint work for $D\rightarrow \pi$ and
$D\rightarrow K$ used an improved staggered quark
action~(``Asqtad'')~\cite{Bernard:2001av} for the light quarks and the
Fermilab action for the heavy quark. To the order implemented in the
study, the Fermilab action corresponds to a re-interpretation of the
clover action. This approach can be used to simulate directly at the
charm and bottom quark mass at the expense of more complicated
discretisation effects. Discretisation errors arising from the final
state energy~($5\%$) and the heavy quark~($7\%$) are estimated to lead
to the largest systematic uncertainties in the calculation~(compared
to the $3\%$ error from the chiral extrapolation). Given the
coarseness of the lattice used, repeating the analysis on a much finer
lattice would enable the estimates of the the discretisation errors to
be confirmed.  Overall, the results are consistent with ours, which
suggests that the systematic effects due to the quenched approximation
are not the dominant source of error.

%
%should I say something about the chiral error?
%

%A technical
%difference in the calculational setup of these works is that the
%Fermilab-MILC-HPQCD study and the~(latter) HPQCD study present results
%from lattices with $a^{-1}\approx 1.6$~GeV, whereas the Fermilab-MILC
%work includes a joint chiral and continuum limit extrapolation.

%Table~\ref{tab:comparison_with_other_works} shows that our results for
%$f_+(0)$ for $D \to \pi$ and $D \to K$ decays are slightly larger than
%those quoted by Ref.~\cite{Aubin:2004ej}. However, this discrepancy is
%not as severe as in the case of Ref.~\cite{Abada:2000ty}: in the worst
%case ($D \to \pi$) it is within $2\sigma$, while for $D \to K$ it is
%within the statistical error uncertainty. It is interesting to observe
%that the results for the $D \to \pi$ decay from the two quenched
%calculations (ours and the APE work) lie on opposite sides of the
%unquenched result, 

For the decay $B \to \pi$, Fermilab and MILC used the same quark
actions as for the study of $D$ decays. Using the 5 light quark masses
at $a=0.12$~fm and 2 light quark masses at $a=0.09$~fm they performed
a joint continuum and chiral extrapolation which removed some of the
discretisation effects. They estimated that a $3\%$ discretisation
error arising from the heavy quark remains after the extrapolation.
The results at finite $q^2$ are compared with ours in
Figure~\ref{fig:beauty_decays_extrapolated_curves}, with statistical
and chiral extrapolation errors only~(which cannot be separated). A
value for $f_+(0)$ is not given in Ref.~\cite{Bailey:2008wp} which
focuses on extracting $|V_{ub}|$ at finite $q^2$ using the
parameterisation of Bourrely, Caprini and
Lellouch~\cite{Bourrely:2008za}. However, an earlier analysis on the
$0.12$~fm lattices only was reported in
Ref.~\cite{Okamoto:2004xg}. Their result for $f_+(0)$ is given
in Table~\ref{tab:comparison_with_other_works}.

HPQCD performed the calculations for the $B\to\pi$ decay on the MILC
configurations using Asqtad light quarks and NRQCD for the $b$
quark. Use of the latter enables direct simulations at the $b$ quark
mass. However, as NRQCD is an effective theory the continuum limit
cannot be taken and scaling in the lattice results must be
demonstrated at finite $a$. Results from the coarse lattice are shown
in Figure~\ref{fig:beauty_decays_extrapolated_curves}, with
statistical and chiral extrapolation errors only and for $f_+(0)$ in
Table~\ref{tab:comparison_with_other_works}. A limited comparison of
results on the finer lattice for one light quark mass did not indicate
that the discretisation errors are large. The systematic errors are
dominated by the estimated $9\%$ uncertainty in the renormalisation
factors which are calculated to 2 loops in perturbation theory.

The Fermilab-MILC and HPQCD results are consistent with each other to
within $2\sigma$ and are also consistent with our results and those of
the APE collaboration. As for the studies of $D$ decays this suggests
that quenching is not the dominant systematic error in the calculation
of $B\to\pi$ decay. Similarly, unquenched results on finer lattices
are needed to investigate the discretisation effects. Finally, note
that in order not to overload
Figures~\ref{fig:charm_decays_extrapolated_curves}
and~\ref{fig:beauty_decays_extrapolated_curves}, we do not show the
(older) quenched results of the Fermilab
group~\cite{ElKhadra:2001rv}. For $B\to\pi$ decays these results are
within the range of the other existing calculations, whereas for
D-decays the form factors come out $10-20\%$ larger compared to most
other calculations and also the new unquenched results obtained with
similar methods.

%within 2 sigma two milc results

%in Table~\ref{tab:comparison_with_other_works}
%we also compare the results with those obtained by the HPQCD
%collaboration in Ref.~\cite{Dalgic:2006dt}: in this case, both
%quenched calculations yield essentially the same result, while the two
%works with $N_f=2+1$ species of dynamical fermions report different
%results (one of which lies above, the other below ours). In
%particular, the number quoted by Ref.~\cite{Dalgic:2006dt} is the
%largest for this decay: it is still compatible with our result within
%one standard deviation, while it differs from the one reported in the
%Fermilab-MILC work~\cite{Okamoto:2004xg} by almost $2 \sigma$.

A different type of comparison can be made with the estimates obtained
in the framework of LCSR. This analytical approach is, to some extent,
complementary to lattice calculations, since it allows one to
calculate the form factors directly at large recoil, albeit with some
assumptions. Figure~\ref{fig:beauty_decays_extrapolated_curves}
compares our extrapolation of the $f^{B \rightarrow \pi}_+(q^2)$ form
factor in the region $q^2 < 12$~GeV$^2$ with the direct LCSR
calculation~\cite{Ball:2004ye,Duplancic:2008ix}. Their predictions are
compatible with our results.  Similar consistency is found between
lattice and LCSR calculations of $f_+(0)$, as seen in
Table~\ref{tab:comparison_with_other_works}, for both $B$ and $D$
decays.
%In particular, for the
%$D$ meson decays we compare our results with
%Ref.~\cite{Khodjamirian:2000ds} and Ref.~\cite{Ball:2006yd}, which
%quote very similar results (although with a relatively large
%uncertainty), in excellent agreement with Ref.~\cite{Aubin:2004ej},
%and compatible with our computation.
%In particular, we note a very
%good agreement between the LCSR prediction, the unquenched results,
%and ours for the $D \to K$ decay. On the other hand, as we discussed
%above, our result for the $D \to \pi$ decay seems to slightly
%overshoot the numbers predicted by LCSR, as well as the unquenched
%simulation result; although the discrepancy is not severe, this effect
%may perhaps be due to the chiral extrapolation of our data, given that
%our lightest pseudoscalar meson states are relatively heavy, at
%approximately $0.5$~GeV.
%As it concerns $B$ and $B_s$ meson decays, in
%Table~\ref{tab:comparison_with_other_works} we also compare our
%results with the LCSR predictions derived in Refs.~\cite{Ball:2004ye,
%  Duplancic:2008ix, Duplancic:2008tk, Wu:2009kq}: in these cases, the
Note that the uncertainty quoted for $f_+(0)$ for $B$ decays is
smaller than that for $D$ meson decays, and comparable with the
precision of the lattice results.  However, while LCSR provides a
systematic approach for calculating these quantities it is by
definition approximate and the errors cannot be reduced below
$10-15\%$, unlike the lattice approach, which is systematically
improveable.

%For
%the $B \to \pi$ and $B \to K$ decays, our results reveal excellent
%agreement with the LCSR predictions, being compatible within less than
%one $\sigma$, while for the $B_s \to K$ form factor the LCSR
%prediction from Ref.~\cite{Duplancic:2008tk} is slightly larger than
%the value we calculated (but still compatible within less than
%$2\sigma$).

\begin{table}[h]
\centering
\begin{tabular}{|c|c||c|c|c|}  \hline
Decay & This work & Other results & Source & Method \\
\hline
\hline
 $ D \rightarrow \pi $ 	& $0.74(6)(4)$ & $0.64(3)(6)$ & Fermilab-MILC-HPQCD~\cite{Aubin:2004ej} & $N_f=2\!+\!1$ LQCD \\
 	 	&  	& $0.57(6)(1)$ & APE~\cite{Abada:2000ty} & $N_f=0$ LQCD \\
 	 	&  	& $0.65(11)$ & Khodjamirian \emph{et al.}~\cite{Khodjamirian:2000ds} & LCSR \\
 	 	&  	& $0.63(11)$ & Ball~\cite{Ball:2006yd} & LCSR \\
\hline
 $ D \rightarrow  K $   & $0.78(5)(4)$  &  $0.73(3)(7)$ & Fermilab-MILC-HPQCD~\cite{Aubin:2004ej} & $N_f=2\!+\!1$ LQCD \\
  		& 	&  $0.66(4)(1)$ & APE~\cite{Abada:2000ty} & $N_f=0$ LQCD \\
  		& 	&   $0.78(11)$ & Khodjamirian \emph{et al.}~\cite{Khodjamirian:2000ds} & LCSR \\
 	 	&  	& $0.75(12)$ & Ball~\cite{Ball:2006yd} & LCSR \\
\hline
 $ D_s \rightarrow K $ 	& $0.68(4)(3)$  &  & & \\
\hline
 $ B \rightarrow \pi $ 	&  $0.27(7)(5) $ & $0.23(2)(3)$ & Fermilab-MILC~\cite{Okamoto:2004xg} & $N_f=2\!+\!1$ LQCD \\
 		 	&  	      & $0.31(5)(4)$ & HPQCD~\cite{Dalgic:2006dt} & $N_f=2\!+\!1$ LQCD \\
		 	&	      & $0.26(5)(4)$ & APE~\cite{Abada:2000ty} & $N_f=0$ LQCD \\
 		 	&             & $0.258(31)$ & Ball and Zwicky~\cite{Ball:2004ye} & LCSR \\
 		 	&             & $0.26(4)$ & Duplan\v ci\'c \emph{et al.}~\cite{Duplancic:2008ix} & LCSR \\
 		 	&             & $0.26(5)$ & Wu and Huang~\cite{Wu:2009kq} & LCSR \\
\hline
 $ B \rightarrow K  $  	&  $0.32(6)(6)$  & $0.331(41)$ & Ball and Zwicky~\cite{Ball:2004ye} & LCSR \\
			&	     & $0.36(5)$ & Duplan\v ci\'c \emph{et al.}~\cite{Duplancic:2008tk} & LCSR \\
 		 	&             & $0.33(8)$ & Wu and Huang~\cite{Wu:2009kq} & LCSR \\
\hline
 $ B_s \rightarrow K $ 	& $0.23(5)(4)$ & $0.30(4)$ & Duplan\v ci\'c \emph{et al.}~\cite{Duplancic:2008tk} & LCSR \\
\hline
\end{tabular}
\caption{Comparison of the results for $f_+(0)$ of the present work
  with other calculations, obtained from lattice QCD (LQCD)
  simulations or from light-cone sum rules (LCSR) by various groups. Where two errors are quoted the first is statistical and the second is the combined systematic errors. }
\label{tab:comparison_with_other_works}
\end{table}

\section{Conclusions}
\label{sec:conclusions}

In this article we have presented a lattice QCD calculation of the form factors associated with semileptonic decays of heavy mesons.

We have performed a quenched calculation on a very fine lattice with
$\beta=6.6$ ($a=0.04$ fm), which allows us to treat the $D$ meson
decays in a fully relativistic setup, and to get close to the region
corresponding to the physical $B$ meson mass. The importance of small
lattice spacings for heavy-quark simulations has recently become clear
in the context of the determination of $f_{D_s}$, the decay constant
of the $D_s$ meson. In spite of $O(a)$ improvement, a continuum
extrapolation linear in $a^2$ seems to be reliable only for lattice
spacings below about 0.07~fm in the quenched
approximation~\cite{AliKhan:2007tm,Heitger:2008jq}. Depending on the
particular improvement condition, even a non-monotonous $a$ dependence
can appear on coarser lattices.

In this work we have investigated to which extent the systematic
effects caused by lattice discretization and long-ranged
extrapolations to the physical heavy meson masses may influence the
results of different lattice calculations in which all other sources
of systematic errors are treated in a similar way. For these reasons,
the results of our study can be directly compared with those by the
APE collaboration in Ref.~\cite{Abada:2000ty}, which reports a very
similar calculation on a coarser lattice at $\beta=6.2$ ($a\simeq
0.07$~fm) with the same lattice action and currents. Adjusting the APE
results so that they comply with our procedure for setting the
physical value of the lattice spacing, we find quite large
discrepancies of roughly $2.5\sigma$~($D\rightarrow \pi$) and
$2\sigma$~($D\rightarrow K$). If we assume that $O(a^2)$ errors are
the dominant source of this effect, the difference in the results of
the two studies suggests an upper limit on the discretization errors
of approximately $0.08$ or slightly above $1\sigma$ in our numbers for
$f_+^{D\rightarrow \pi}(0)$ and $0.23$ or $3-4\sigma$ in the APE
results.

It is, however, to be noted that the interpretation of this difference
as a mere discretization error is somewhat more ambiguous than in the
case of the decay constants considered
in~\cite{AliKhan:2007tm,Heitger:2008jq}, because the momentum transfer
$q^2$ adds another parameter that has to be adjusted before the
comparison can be attempted. The corresponding comparison for $B$
decays can, in addition, be undermined by the long-ranged
extrapolations in the heavy quark mass and/or $q^2$. These results
suggest that, for high-precision phenomenological applications,
completely reliable relativistic lattice calculations of these form
factors could require even finer spacings, and that, for dynamical
simulations at realistic pion masses, this goal might be difficult to
achieve in the near future. While we believe that the progress in
computational power will eventually allow one to realize this
formidable task, it is fair to say that, for the moment, the less
demanding approaches which interpolate between the $D$ meson scale and
non-relativistic results provide a valid alternative.

Finally, a few words are in order about the general perspective for
calculations of the semileptonic form factors of heavy mesons. Form
factors of $B$ decays at small values of the relativistic momentum
transfer $q^2$ involve a light meson with momentum up to 2.5~GeV in
the final state and are very difficult to calculate on the lattice,
mainly because no lattice effective field theory formulation is known
for this kinematics that would allow for the consistent separation of
the large scales of the order of the heavy quark mass, as implemented
in the Soft-Collinear Effective Theory.

Thus one is left with two choices. The first one is to calculate the
form factors at moderate recoil ($m_B^2-q^2\sim
O(m_B\Lambda_{\mbox{\scriptsize{QCD}}})$) using, e.g., the HQET or
NRQCD expansion and then to extrapolate to large recoil
($m_B^2-q^2\sim O(m^2_B)$) guided by the dispersion relations. The
advantage of this approach is that the calculations can be performed
on relatively coarse and thus not very large (in lattice units)
lattices. Therefore dynamical fermions may be included, high
statistical accuracy can be achieved as well as a better control over
the chiral extrapolation. The disadvantage is that a reliable
extrapolation from the $q^2 > 12-15$~GeV$^2$ regime accessible in this
method to $q^2=0$ may be subtle. However, this problem may be
alleviated by a promising new approach, ``moving
NRQCD''~\cite{Horgan:2009ti}, which formulates NRQCD in a reference
frame where the heavy quark is moving with a velocity $v$. By giving
the $B$ meson significant spatial momentum, relatively low $q^2$ can
be achieved for lower values of the final state momentum thus avoiding
large discretisation effects.

{} For the particular case of the $B\to\pi$ semileptonic decay the
problem of simulating at large recoil can be avoided, at least in
principle, since the shape of the form factor $f_+(q^2)$ can be
extracted from the experimental data on the partial branching fraction
in different $q^2$ bins, see, e.g., Ref.~\cite{Aubert:2006px}. The
normalization can then be fixed by comparison to lattice data in the
$q^2\sim 10-20$~GeV$^2$ range. This strategy (see
Ref.~\cite{Flynn:2007ii} for a detailed discussion) is indeed
promising and may lead to a considerable improvement in the accuracy
of the $|V_{ub}|$ determination from exclusive $B$ decays provided the
combined data analysis using the full statistics of the BaBar and
Belle experiments ($\sim 4\cdot 10^8~\bar B B$ pairs) becomes
available. However, for rare decays, such as $B\to K^*\gamma$, $B\to
K^*\mu^+\mu^-$ etc., which are likely to take the central stage at
LHCb and super-$B$ factories, a similar strategy seems to be
unfeasible.

The second choice are simulations with fully relativistic heavy quarks
on very fine and large (in lattice units) lattices. This procedure is
presently bound to the quenched approximation, but the benefit is that
the extrapolation in the heavy quark mass and potentially also in
$q^2$ is of much shorter range. It goes without saying that the
inclusion of dynamical fermions and the approach to the chiral limit
are also important problems, but presently it is impossible to address
all relevant issues within one calculation.

In our opinion both methods are justified and we have chosen the
second option in this paper. It turns out that our final results for,
e.g., the $B \rightarrow \pi$ decays are consistent with
determinations based on dynamical simulations and LCSR. This may
indicate that the quenching effects are rather moderate. From our
experience, the main problem that limits the usefulness of this
approach is the construction of sources for the light hadrons which
yield a good overlap with states of large momentum. It seems that the
presently used sources are not good enough in this respect. Improved
sources have to be developed if a similar calculation is attempted on
a larger scale in the future.

\section*{Acknowledgements}

We warmly thank Tommy~Burch for collaboration in the early stages of this project.
The numerical calculations were performed on the Hitachi SR8000 at LRZ Munich.
This work was supported by DFG (Forschergruppe Gitter-Hadronen-Ph\"anomenologie) and GSI.
A.A.K. thanks the DFG and ``Berliner Programm zur F\"orderung der Chancengleichheit f\"ur Frauen in Forschung und Lehre''
for financial support.
S.C. acknowledges financial support from the
Claussen-Simon-Foundation (Stifterverband f\"ur die Deutsche Wissenschaft).
M.P. gratefully acknowledges support from the Alexander~von~Humboldt Stiftung/Foundation and from INFN.
The University of Regensburg hosts the Collaborative Research Center SFB/TR 55 ``Hadron Physics from Lattice QCD''.

\appendix{}
\section{Simulation results}
\renewcommand{\theequation}{A.\arabic{equation}}
\setcounter{equation}{0}

\begin{figure}
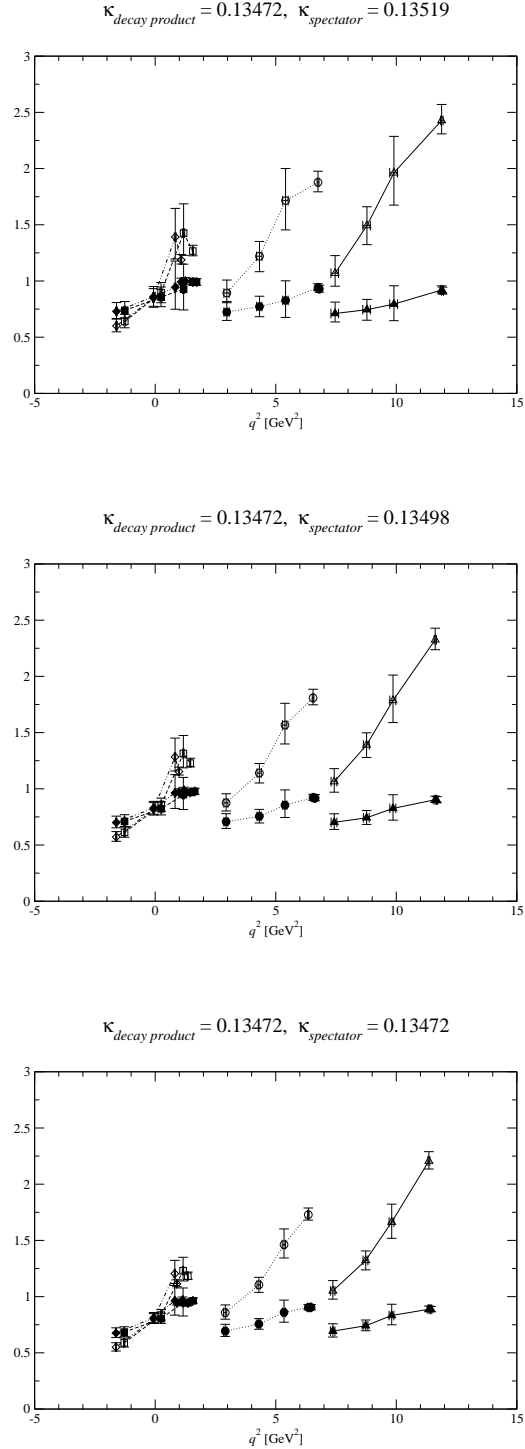

\centerline{
\includegraphics[width=.40\textwidth]{simulated_points_in_GeVsq_with_lines_from_kdecayproduct_13472_kspectator_13519.eps}
}
\vspace{10mm}
\centerline{
\includegraphics[width=.40\textwidth]{simulated_points_in_GeVsq_with_lines_from_kdecayproduct_13472_kspectator_13498.eps}
}
\vspace{10mm}
\centerline{
\includegraphics[width=.40\textwidth]{simulated_points_in_GeVsq_with_lines_from_kdecayproduct_13472_kspectator_13472.eps}
}
  \caption{Sample of form factors directly measured in our
    simulations. The three panels show the results obtained for
    $\kappa_{decay \; product}=0.13472$, and for different values of
    $\kappa_{spectator}$. In each plot, the results for $f_0$ (denoted
    by full symbols) and for $f_+$ (empty symbols) are plotted against
    the square of the transferred momentum $q^2$. The results for
    different values of $\kappa_{heavy}$ are displayed using different
    symbols: diamonds ($\kappa_{heavy}=0.13$), squares
    ($\kappa_{heavy}=0.129$), circles ($\kappa_{heavy}=0.121$) and
    triangles ($\kappa_{heavy}=0.115$).}
  \label{fig:simulated_points_kdecayproduct_13472}
\end{figure}

Fig.~\ref{fig:simulated_points_kdecayproduct_13472} shows a subset of
the form factors $f_0(q^2)$ and $f_+(q^2)$ that we extracted from our
simulations, for different combinations of the $\kappa$ values for the
heavy and spectator quarks, with $\kappa_{decay\; product}=0.13472$.
In Table~\ref{tab:unextrapolated_BK_fit_results} we present our
results for the fits to the simulation data with the
Be\'cirevi\'c--Kaidalov parameterization~\cite{Becirevic:1999kt}
according to
Eq.~(\protect\ref{eq:Becirevic_Kaidalov_parametrization}).  This
parameterization uses the vector meson mass $m_{H^\star}$. Our results for $m_{H^\star}$ in lattice units are shown in Table~\ref{tab:hstar_masses}.

\begin{table}[!h]
\centering
\begin{tabular}{|c|c|c|c|c|c|c|}  \hline
$\kappa_{dec.~prod.}$ & $\kappa_{heavy}$ & $\kappa_{spect.}$ &  $c_{BK} \cdot (1-\alpha)$ & $1/\beta$ & $\alpha$ & $\chi^2/$d.o.f.\\
\hline
 $ 0.13519$ & $0.13$ & $0.13519 $ & $ 0.775^{+39}_{-45} $ & $ 0.580^{+85}_{-79} $ & $ 0.01^{+11}_{-10} $ & $  0.77 $ \\
 $ 0.13519$ & $0.129$ & $0.13519 $ & $ 0.724^{+49}_{-47} $ & $ 0.604^{+95}_{-90} $ & $ 0.09^{+12}_{-12} $ & $  1.08 $ \\
 $ 0.13519$ & $0.121$ & $0.13519 $ & $ 0.484^{+88}_{-72} $ & $ 0.73^{+11}_{-12} $ & $ 0.49^{+15}_{-16} $ & $  0.66 $ \\
 $ 0.13519$ & $0.115$ & $0.13519 $ & $ 0.39^{+10}_{-8} $ & $ 0.78^{+11}_{-14} $ & $ 0.66^{+13}_{-16} $ & $  0.45 $ \\
 $ 0.13519$ & $0.13$ & $0.13498 $ & $ 0.742^{+32}_{-37} $ & $ 0.658^{+79}_{-75} $ & $ 0.229^{+88}_{-81} $ & $  1.35 $ \\
 $ 0.13519$ & $0.129$ & $0.13498 $ & $ 0.674^{+37}_{-35} $ & $ 0.714^{+81}_{-85} $ & $ 0.337^{+91}_{-91} $ & $  1.51 $ \\
 $ 0.13519$ & $0.121$ & $0.13498 $ & $ 0.442^{+61}_{-52} $ & $ 0.808^{+87}_{-96} $ & $ 0.65^{+11}_{-11} $ & $  0.71 $ \\
 $ 0.13519$ & $0.115$ & $0.13498 $ & $ 0.364^{+68}_{-52} $ & $ 0.84^{+8}_{-10} $ & $ 0.76^{+9}_{-11} $ & $  0.36 $ \\
 $ 0.13519$ & $0.13$ & $0.13472 $ & $ 0.750^{+29}_{-30} $ & $ 0.670^{+77}_{-71} $ & $ 0.338^{+75}_{-74} $ & $  1.05 $ \\
 $ 0.13519$ & $0.129$ & $0.13472 $ & $ 0.685^{+34}_{-29} $ & $ 0.730^{+70}_{-82} $ & $ 0.437^{+76}_{-77} $ & $  0.99 $ \\
 $ 0.13519$ & $0.121$ & $0.13472 $ & $ 0.456^{+49}_{-42} $ & $ 0.821^{+78}_{-89} $ & $ 0.709^{+86}_{-98} $ & $  0.39 $ \\
 $ 0.13519$ & $0.115$ & $0.13472 $ & $ 0.387^{+59}_{-46} $ & $ 0.829^{+76}_{-88} $ & $ 0.787^{+78}_{-98} $ & $  0.21 $ \\
 $ 0.13498$ & $0.13$ & $0.13519 $ & $ 0.783^{+33}_{-35} $ & $ 0.587^{+78}_{-68} $ & $ 0.065^{+94}_{-83} $ & $  0.87 $ \\
 $ 0.13498$ & $0.129$ & $0.13519 $ & $ 0.740^{+41}_{-38} $ & $ 0.596^{+85}_{-70} $ & $ 0.12^{+10}_{-9} $ & $  1.20 $ \\
 $ 0.13498$ & $0.121$ & $0.13519 $ & $ 0.536^{+71}_{-65} $ & $ 0.67^{+11}_{-12} $ & $ 0.44^{+14}_{-14} $ & $  0.78 $ \\
 $ 0.13498$ & $0.115$ & $0.13519 $ & $ 0.419^{+84}_{-67} $ & $ 0.76^{+10}_{-12} $ & $ 0.64^{+12}_{-15} $ & $  0.61 $ \\
 $ 0.13498$ & $0.13$ & $0.13498 $ & $ 0.781^{+30}_{-28} $ & $ 0.613^{+63}_{-61} $ & $ 0.200^{+67}_{-66} $ & $  1.03 $ \\
 $ 0.13498$ & $0.129$ & $0.13498 $ & $ 0.735^{+34}_{-29} $ & $ 0.627^{+67}_{-66} $ & $ 0.256^{+71}_{-72} $ & $  1.24 $ \\
 $ 0.13498$ & $0.121$ & $0.13498 $ & $ 0.512^{+54}_{-46} $ & $ 0.723^{+77}_{-91} $ & $ 0.557^{+92}_{-98} $ & $  0.56 $ \\
 $ 0.13498$ & $0.115$ & $0.13498 $ & $ 0.428^{+60}_{-47} $ & $ 0.765^{+77}_{-97} $ & $ 0.69^{+9}_{-11} $ & $  0.36 $ \\
 $ 0.13498$ & $0.13$ & $0.13472 $ & $ 0.779^{+28}_{-26} $ & $ 0.664^{+62}_{-61} $ & $ 0.325^{+62}_{-62} $ & $  1.18 $ \\
 $ 0.13498$ & $0.129$ & $0.13472 $ & $ 0.726^{+27}_{-27} $ & $ 0.690^{+61}_{-63} $ & $ 0.386^{+62}_{-65} $ & $  1.03 $ \\
 $ 0.13498$ & $0.121$ & $0.13472 $ & $ 0.484^{+41}_{-36} $ & $ 0.803^{+65}_{-72} $ & $ 0.675^{+73}_{-85} $ & $  0.48 $ \\
 $ 0.13498$ & $0.115$ & $0.13472 $ & $ 0.407^{+46}_{-38} $ & $ 0.822^{+65}_{-75} $ & $ 0.766^{+70}_{-83} $ & $  0.54 $ \\
 $ 0.13472$ & $0.13$ & $0.13519 $ & $ 0.821^{+32}_{-31} $ & $ 0.565^{+74}_{-61} $ & $ 0.053^{+80}_{-70} $ & $  0.82 $ \\
 $ 0.13472$ & $0.129$ & $0.13519 $ & $ 0.781^{+34}_{-33} $ & $ 0.560^{+78}_{-65} $ & $ 0.094^{+90}_{-77} $ & $  0.97 $ \\
 $ 0.13472$ & $0.121$ & $0.13519 $ & $ 0.565^{+60}_{-54} $ & $ 0.66^{+11}_{-10} $ & $ 0.42^{+14}_{-12} $ & $  0.80 $ \\
 $ 0.13472$ & $0.115$ & $0.13519 $ & $ 0.472^{+78}_{-62} $ & $ 0.72^{+11}_{-12} $ & $ 0.59^{+13}_{-13} $ & $  0.69 $ \\
 $ 0.13472$ & $0.13$ & $0.13498 $ & $ 0.802^{+28}_{-26} $ & $ 0.620^{+60}_{-59} $ & $ 0.197^{+62}_{-58} $ & $  0.86 $ \\
 $ 0.13472$ & $0.129$ & $0.13498 $ & $ 0.762^{+30}_{-27} $ & $ 0.625^{+63}_{-59} $ & $ 0.244^{+68}_{-65} $ & $  1.15 $ \\
 $ 0.13472$ & $0.121$ & $0.13498 $ & $ 0.530^{+43}_{-42} $ & $ 0.734^{+78}_{-76} $ & $ 0.549^{+95}_{-88} $ & $  0.68 $ \\
 $ 0.13472$ & $0.115$ & $0.13498 $ & $ 0.450^{+50}_{-46} $ & $ 0.764^{+77}_{-83} $ & $ 0.665^{+89}_{-92} $ & $  0.59 $ \\
 $ 0.13472$ & $0.13$ & $0.13472 $ & $ 0.803^{+26}_{-22} $ & $ 0.648^{+50}_{-53} $ & $ 0.290^{+52}_{-53} $ & $  1.01 $ \\
 $ 0.13472$ & $0.129$ & $0.13472 $ & $ 0.751^{+27}_{-23} $ & $ 0.673^{+53}_{-53} $ & $ 0.353^{+58}_{-59} $ & $  0.80 $ \\
 $ 0.13472$ & $0.121$ & $0.13472 $ & $ 0.524^{+35}_{-31} $ & $ 0.768^{+57}_{-66} $ & $ 0.616^{+69}_{-71} $ & $  0.79 $ \\
 $ 0.13472$ & $0.115$ & $0.13472 $ & $ 0.425^{+39}_{-33} $ & $ 0.823^{+58}_{-67} $ & $ 0.749^{+65}_{-75} $ & $  1.07 $ \\
\hline
\end{tabular}
\caption{Results of the fits of the lattice data with Eq.~(\protect\ref{eq:Becirevic_Kaidalov_parametrization}). Note that at vanishing $q^2$ one has $f_+(0)=f_0(0)$, which is given by $c_{BK} \cdot (1-\alpha)$ (fourth column of this Table).} \label{tab:unextrapolated_BK_fit_results}
\end{table}

\begin{table}[!h]
\centering
\phantom{-------}
\begin{tabular}{|c|c|c|c|} \hline
$\kappa_{heavy}$ & $\kappa_{light}$ & $am_{H}$  & $am_{H^\star}$ \\
\hline
 $  0.13$ & $ 0.13519$ & $0.3681^{+13}_{-11}$& $  0.3949^{+18}_{-16}  $ \\
 $  0.13$ & $ 0.13498$ & $0.3762^{+13}_{-10}$& $  0.4017^{+16}_{-14}  $ \\
 $  0.13$ & $ 0.13472$ &  $0.3867^{+11}_{-10}$& $  0.4110^{+14}_{-14}  $ \\
 $ 0.129$ & $ 0.13519$ & $0.4060^{+13}_{-11}$& $  0.4300^{+17}_{-15}  $ \\
 $ 0.129$ & $ 0.13498$ & $0.4138^{+13}_{-11}$& $  0.4366^{+16}_{-15}  $ \\
 $ 0.129$ & $ 0.13472$ & $0.4240^{+12}_{-10}$& $  0.4458^{+14}_{-14}  $ \\
 $ 0.121$ & $ 0.13519$ & $0.6672^{+16}_{-13}$& $  0.6804^{+20}_{-17}  $ \\
 $ 0.121$ & $ 0.13498$ & $0.6743^{+14}_{-12}$& $  0.6868^{+17}_{-16}  $ \\
 $ 0.121$ & $ 0.13472$ & $0.6836^{+12}_{-12}$& $  0.6956^{+15}_{-14}  $ \\
 $ 0.115$ & $ 0.13519$ & $0.8369^{+17}_{-14}$& $  0.8460^{+20}_{-18}  $ \\
 $ 0.115$ & $ 0.13498$ &  $0.8437^{+15}_{-13}$& $  0.8523^{+17}_{-16}  $ \\
 $ 0.115$ & $ 0.13472$ & $0.8527^{+13}_{-12}$& $  0.8611^{+15}_{-14}  $ \\
\hline
\end{tabular}
\phantom{-------}
\caption{Masses of the heavy-light pseudoscalar and vector states, $H$ and $H^\star$, respectively, in lattice units, for the different $(\kappa_{heavy},\kappa_{light} )$ combinations.} \label{tab:hstar_masses}
\end{table}

\begin{figure}
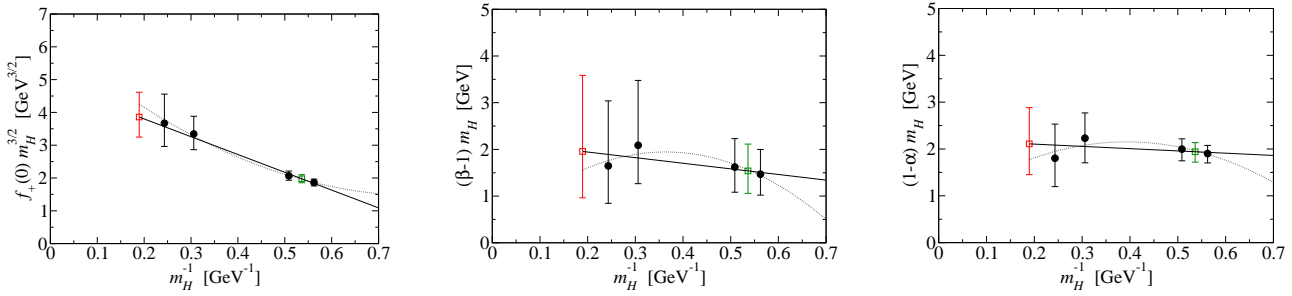

  \centerline{\includegraphics[width=.30\textwidth]{fplusatzero_times_M_H_to_three_halves_decay_from_H_to_K.eps}\hfill \includegraphics[width=.30\textwidth]{beta_minus_one_times_M_H_decay_from_H_to_K.eps}\hfill\includegraphics[width=.30\textwidth]{one_minus_alpha_times_M_H_decay_from_H_to_K.eps}}
  \caption{Extrapolation and interpolation of the (chirally extrapolated) Be\'cirevi\'c--Kaidalov parameters in the heavy quark mass. The plots show the results obtained for the combinations $f_+(0)\cdot m_H^{3/2}$ (l.h.s. panel), $(\beta-1) m_H$ (central panel) and $(1-\alpha) m_H$ (r.h.s. panel), for decays of a $B$ (red squares) or of a $D$ meson (green squares) to a kaon. The results are obtained using linear fits in $m_H^{-1}$ (solid lines) according to     Eq.~(\protect\ref{eq:linear_heavy_quark_expansion}); for comparison, the curves resulting from fits to quadratic order in $m_H^{-1}$ (dotted lines) are also shown.}
  \label{fig:heavy_extrapolation_of_all_BK_parameters_for_H_to_K}
\end{figure}

The parameters $f_+(0)$, $\alpha$ and $\beta$ are then extrapolated to
the light-quark mass as described in
Section~\ref{sec:physical_results} and illustrated in
Fig.~\ref{fig:pion_chiral_extrapolation_kheavy_11500}.  Finally, the
parameters describing the physical form factors are obtained through
extrapolation of these values to the physical $B$ meson mass (or
through interpolation to the physical $D$ meson mass), according to
the heavy-quark scaling laws, see
Fig.~\ref{fig:heavy_extrapolation_of_all_BK_parameters_for_H_to_K}.

\begin{thebibliography}{99}

%\cite{Amsler:2008zz}
\bibitem{Amsler:2008zz}
  R.~Kowalewski and T.~Mannel,
  %{\it Determination of $V_{cb}$ and $V_{ub}$,}
  in:
  C.~Amsler {\it et al.}  [Particle Data Group],
  {\it Review of particle physics,}
  \plb{667}{2008}{1}.
  %%CITATION = PHLTA,B667,1;%%

%\cite{Buchalla:2008jp}
\bibitem{Buchalla:2008jp}
  M.~Artuso {\it et al.},
  %{\it B, D and K decays,}
  arXiv:0801.1833 [hep-ph].
  %%CITATION = ARXIV:0801.1833;%%

%\cite{Beneke:1999br}
\bibitem{Beneke:1999br}
  M.~Beneke, G.~Buchalla, M.~Neubert and C.~T.~Sachrajda,
  %{\it {QCD} factorization for B $\rightarrow \pi \pi$ decays: Strong phases and CP  violation in the heavy quark limit,}
  \prl{83}{1999}{1914}
  [arXiv:hep-ph/9905312].
  %%CITATION = PRLTA,83,1914;%%

%\cite{Beneke:2001ev}
\bibitem{Beneke:2001ev}
  M.~Beneke, G.~Buchalla, M.~Neubert and C.~T.~Sachrajda,
  %{\it QCD factorization in B $\rightarrow \pi K$, $\pi \pi$ decays and extraction of Wolfenstein parameters,}
  \npb{606}{2001}{245}
  [arXiv:hep-ph/0104110].
  %%CITATION = NUPHA,B606,245;%%

%\cite{Aubin:2004ej}
\bibitem{Aubin:2004ej}
  C.~Aubin {\it et al.}  [Fermilab Lattice Collaboration and MILC Collaboration and HPQCD Collaboration],
  %{\it Semileptonic decays of D mesons in three-flavor lattice QCD,}
  \prl{94}{2005}{011601}
  [arXiv:hep-ph/0408306].
  %%CITATION = PRLTA,94,011601;%%

%\cite{Okamoto:2005zg}
\bibitem{Okamoto:2005zg}
  M.~Okamoto,
  %{\it Full determination of the CKM matrix using recent results from lattice QCD,}
  PoS {\bf LAT2005} (2006) 013
  [arXiv:hep-lat/0510113].
  %%CITATION = POSCI,LAT2005,013;%%


%\cite{Dalgic:2006dt}
\bibitem{Dalgic:2006dt}
  E.~Dalgic, A.~Gray, M.~Wingate, C.~T.~H.~Davies, G.~P.~Lepage and J.~Shigemitsu,
  %{\it B Meson Semileptonic Form Factors from Unquenched Lattice QCD,}
  \prd{73}{2006}{074502}
  [Erratum-ibid. {\bf D 75} (2007) 119906]
  [arXiv:hep-lat/0601021].
  %%CITATION = PHRVA,D73,074502;%%


%\cite{Becirevic:1999kt}
\bibitem{Becirevic:1999kt}
  D.~Be\'cirevi\'c and A.~B.~Kaidalov,
  %{\it Comment on the heavy $\rightarrow$ light form factors,}
  \plb{478}{2000}{417}
  [arXiv:hep-ph/9904490].
  %%CITATION = PHLTA,B478,417;%%


%\cite{Albertus:2005ud}
\bibitem{Albertus:2005ud}
  C.~Albertus, J.~M.~Flynn, E.~Hern\'andez, J.~Nieves and J.~M.~Verde-Velasco,
  %{\it Semileptonic B $\rightarrow \pi$ decays from an Omnes improved nonrelativistic constituent quark model,}
  \prd{72}{2005}{033002}
  [arXiv:hep-ph/0506048].
  %%CITATION = PHRVA,D72,033002;%%

%\cite{Boyd:1994tt}
\bibitem{Boyd:1994tt}
  C.~G.~Boyd, B.~Grinstein and R.~F.~Lebed,
  %{\it Constraints On Form-Factors For Exclusive Semileptonic Heavy To Light Meson Decays,}
  \prl{74}{1995}{4603}
  [arXiv:hep-ph/9412324].
  %%CITATION = PRLTA,74,4603;%%

%\cite{DescotesGenon:2008hh}
\bibitem{DescotesGenon:2008hh}
  S.~Descotes-Genon and A.~Le~Yaouanc,
  %{\it Parametrisations of the D $\rightarrow K \ell \nu$ form factor and the determination of $\hat{g}$,}
  \jphg{35}{2008}{115005}
  [arXiv:0804.0203 [hep-ph]].
  %%CITATION = JPHGB,G35,115005;%%

%\cite{Bourrely:2008za}
\bibitem{Bourrely:2008za}
  C.~Bourrely, I.~Caprini and L.~Lellouch,
  %{\it Model-independent description of $B\to \pi l\nu$ decays and a determination of $|V_{ub}|$,}
  arXiv:0807.2722 [hep-ph].
  %%CITATION = ARXIV:0807.2722;%%

%\cite{Balitsky:1989ry}
\bibitem{Balitsky:1989ry}
  I.~I.~Balitsky, V.~M.~Braun and A.~V.~Kolesnichenko,
  %{\it Radiative Decay $\Sigma^+ \to p \gamma$ in Quantum Chromodynamics,}
  \npb{312}{1989}{509}.
  %%CITATION = NUPHA,B312,509;%%

%\cite{Chernyak:1990ag}
\bibitem{Chernyak:1990ag}
  V.~L.~Chernyak and I.~R.~Zhitnitsky,
  %{\it B Meson Exclusive Decays Into Baryons,}
  \npb{345}{1990}{137}.
  %%CITATION = NUPHA,B345,137;%%

%\cite{Ball:2004ye}
\bibitem{Ball:2004ye}
  P.~Ball and R.~Zwicky,
  %{\it New results on B $\rightarrow \pi$, $K$, $\eta$ decay formfactors from light-cone sum rules,}
  \prd{71}{2005}{014015}
  [arXiv:hep-ph/0406232].
  %%CITATION = PHRVA,D71,014015;%%

%\cite{Duplancic:2008ix}
\bibitem{Duplancic:2008ix}
  G.~Duplan\v ci\'c, A.~Khodjamirian, T.~Mannel, B.~Meli\'c and N.~Offen,
  %{\it Light-cone sum rules for $B \to \pi$ form factors revisited,}
  \jhep{04}{2008}{014}
  [arXiv:0801.1796 [hep-ph]].
  %%CITATION = JHEPA,0804,014;%%

%\cite{Duplancic:2008tk}
\bibitem{Duplancic:2008tk}
  G.~Duplan\v ci\'c and B.~Meli\'c,
  %{\it $B$, $B_s \rightarrow K$ form factors: an update of light-cone sum rule results,}
  arXiv:0805.4170 [hep-ph].
  %%CITATION = ARXIV:0805.4170;%%

%\cite{Wu:2009kq}
\bibitem{Wu:2009kq}
  X.~G.~Wu and T.~Huang,
  %{\it Radiative Corrections on the $B\to P$ Form Factors with Chiral Current in the Light-Cone Sum Rules,}
  \prd{79}{2009}{034013}
  [arXiv:0901.2636 [hep-ph]].
  %%CITATION = PHRVA,D79,034013;%%

%\cite{AliKhan:2007tm}
\bibitem{AliKhan:2007tm}
  A.~Ali~Khan, V.~Braun, T.~Burch, M.~G\"ockeler, G.~Lacagnina, A.~Sch\"afer and G.~Schierholz,
  %{\it Decay constants of charm and beauty pseudoscalar heavy-light mesons on fine lattices,}
  \plb{652}{2007}{150}
  [arXiv:hep-lat/0701015].
  %%CITATION = PHLTA,B652,150;%%


%\cite{Heitger:2008jq}
\bibitem{Heitger:2008jq}
  J.~Heitger and A.~J\"uttner,
  %{\it Lattice cutoff effects for $F_{D_s}$ with improved Wilson fermions - a final lesson from the quenched case,}
  arXiv:0812.2200 [hep-lat].
  %%CITATION = ARXIV:0812.2200;%%

%\cite{Narison:2008bc}
\bibitem{Narison:2008bc}
  S.~Narison,
  %{\it $|V_{cd}|$, $|V_{cs}|$ and $f_{D_s}$ from (semi) leptonic $D_{(s)}$-decays : signals of New Physics ?,}
  arXiv:0807.2830 [hep-ph].
  %%CITATION = ARXIV:0807.2830;%%

%\cite{Abada:2000ty}
\bibitem{Abada:2000ty}
  A.~Abada, D.~Be\'cirevi\'c, P.~Boucaud, J.~P.~Leroy, V.~Lubicz and F.~Mescia,
  %{\it Heavy $\rightarrow$ light semileptonic decays of pseudoscalar mesons from lattice QCD,}
  \npb{619}{2001}{565}
  [arXiv:hep-lat/0011065].
  %%CITATION = NUPHA,B619,565;%%

%\cite{Khan:2007hn}
\bibitem{Khan:2007hn}
  A.~Ali~Khan, V.~Braun, T.~Burch, M.~G\"ockeler, G.~Lacagnina, A.~Sch\"afer and G.~Schierholz,
  %{\it Decays of mesons with charm quarks on the lattice,}
  PoS {\bf LAT2007} (2007) 343
  [arXiv:0710.1070 [hep-lat]].
  %%CITATION = POSCI,LAT2007,343;%%

%\cite{Khan:2009eq}
\bibitem{Khan:2009eq}
  A.~Ali~Khan {\it et al.},
  %{\it Matrix elements of heavy-light mesons from a fine lattice,}
  PoS {\bf Confinement8} (2009) 167
  [arXiv:0901.0822 [hep-lat]].
  %%CITATION = ARXIV:0901.0822;%%

%\cite{Eichten:1989zv}
\bibitem{Eichten:1989zv}
  E.~Eichten and B.~R.~Hill,
  %{\it An Effective Field Theory for the Calculation of Matrix Elements Involving Heavy Quarks,}
  \plb{234}{1990}{511}.
  %%CITATION = PHLTA,B234,511;%%

%\cite{Lepage:1992tx}
\bibitem{Lepage:1992tx}
  G.~P.~Lepage, L.~Magnea, C.~Nakhleh, U.~Magnea and K.~Hornbostel,
  %{\it Improved nonrelativistic QCD for heavy quark physics,}
  \prd{46}{1992}{4052}
  [arXiv:hep-lat/9205007].
  %%CITATION = PHRVA,D46,4052;%%

%\cite{Jansen:2008si}
\bibitem{Jansen:2008si}
  K.~Jansen, C.~Michael, A.~Shindler and M.~Wagner  [ETM Collaboration],
  %{\it The static-light meson spectrum from twisted mass lattice QCD,}
  arXiv:0810.1843 [hep-lat].
  %%CITATION = ARXIV:0810.1843;%%

%\cite{DellaMorte:2007ij}
\bibitem{DellaMorte:2007ij}
  M.~Della Morte, S.~D\"urr, D.~Guazzini, R.~Sommer, J.~Heitger and A.~J\"uttner,
  %{\it Heavy-strange meson decay constants in the continuum limit of quenched QCD,}
  \jhep{02}{2008}{078}
  [arXiv:0710.2201 [hep-lat]].
  %%CITATION = JHEPA,0802,078;%%

%\cite{Hein:2000qu}
\bibitem{Hein:2000qu}
  J.~Hein {\it et al.},
  %{\it Scaling of the B and D meson spectrum in lattice QCD,}
  \prd{62}{2000}{074503}
  [arXiv:hep-ph/0003130].
  %%CITATION = PHRVA,D62,074503;%%

%\cite{AliKhan:2001jg}
\bibitem{AliKhan:2001jg}
  A.~Ali Khan {\it et al.}  [CP-PACS Collaboration],
  %{\it B meson decay constant from two-flavor lattice QCD with non-relativistic heavy quarks,}
  \prd{64}{2001}{054504}
  [arXiv:hep-lat/0103020].
  %%CITATION = PHRVA,D64,054504;%%

%\cite{ElKhadra:1996mp}
\bibitem{ElKhadra:1996mp}
  A.~X.~El-Khadra, A.~S.~Kronfeld and P.~B.~Mackenzie,
  %{\it Massive Fermions in Lattice Gauge Theory,}
  \prd{55}{1997}{3933}
  [arXiv:hep-lat/9604004].
  %%CITATION = PHRVA,D55,3933;%%

%\cite{Heitger:2003xg}
\bibitem{Heitger:2003xg}
  J.~Heitger, M.~Kurth and R.~Sommer  [ALPHA Collaboration],
  %{\it Non-perturbative renormalization of the static axial current in  quenched QCD,}
  \npb{669}{2003}{173}
  [arXiv:hep-lat/0302019].
  %%CITATION = NUPHA,B669,173;%%

%\cite{DellaMorte:2006sv}
\bibitem{DellaMorte:2006sv}
  M.~Della Morte, P.~Fritzsch and J.~Heitger,
  %{\it Non-perturbative renormalization of the static axial current in two-flavour QCD,}
  \jhep{02}{2007}{079}
  [arXiv:hep-lat/0611036].
  %%CITATION = JHEPA,0702,079;%%

%\cite{Necco:2001xg}
\bibitem{Necco:2001xg}
  S.~Necco and R.~Sommer,
  %{\it The $N_f = 0$ heavy quark potential from short to intermediate  distances,}
  \npb{622}{2002}{328}
  [arXiv:hep-lat/0108008].
  %%CITATION = NUPHA,B622,328;%%

%\cite{Sheikholeslami:1985ij}
\bibitem{Sheikholeslami:1985ij}
  B.~Sheikholeslami and R.~Wohlert,
  %{\it Improved Continuum Limit Lattice Action For QCD With Wilson Fermions,}
  \npb{259}{1985}{572}.
  %%CITATION = NUPHA,B259,572;%%

%\cite{Luscher:1996ug}
\bibitem{Luscher:1996ug}
  M.~L\"uscher, S.~Sint, R.~Sommer, P.~Weisz and U.~Wolff,
  %{\it Non-perturbative O(a) improvement of lattice QCD,}
  \npb{491}{1997}{323}
  [arXiv:hep-lat/9609035].
  %%CITATION = NUPHA,B491,323;%%

%\cite{Luscher:1996jn}
\bibitem{Luscher:1996jn}
  M.~L\"uscher, S.~Sint, R.~Sommer and H.~Wittig,
  %{\it Non-perturbative determination of the axial current normalization  constant in O(a) improved lattice QCD,}
  \npb{491}{1997}{344}
  [arXiv:hep-lat/9611015].
  %%CITATION = NUPHA,B491,344;%%

%\cite{Bakeyev:2003ff}
\bibitem{Bakeyev:2003ff}
  T.~Bakeyev, M.~G\"ockeler, R.~Horsley, D.~Pleiter, P.~E.~L.~Rakow, G.~Schierholz and H.~St\"uben [QCDSF-UKQCD Collaboration],
  %{\it Non-perturbative renormalisation and improvement of the local vector current for quenched and unquenched Wilson fermions,}
  \plb{580}{2004}{197}
  [arXiv:hep-lat/0305014].
  %%CITATION = PHLTA,B580,197;%%

%\cite{Guagnelli:1997db}
\bibitem{Guagnelli:1997db}
  M.~Guagnelli and R.~Sommer,
  %{\it Non-perturbative O(a) improvement of the vector current,}
  \npps{63}{1998}{886}
  [arXiv:hep-lat/9709088].
  %% CITATION = NUPHZ,63,886;%%}

%\cite{Pleiter_thesis}
\bibitem{Pleiter_thesis}
  D.~Pleiter, 
  Ph.D. thesis, 
  Freie Universit\"at Berlin (2000).

%\cite{Charles:1998dr}
\bibitem{Charles:1998dr}
  J.~Charles, A.~Le Yaouanc, L.~Oliver, O.~P\`ene and J.~C.~Raynal,
  %{\it Heavy-to-light form factors in the heavy mass to large energy limit of QCD,}
  \prd{60}{1999}{014001}
  [arXiv:hep-ph/9812358].
  %%CITATION = PHRVA,D60,014001;%%

%\cite{Ball:2006jz}
\bibitem{Ball:2006jz}
  P.~Ball,
  %{\it $|V_{ub}|$ from UTangles and B $\rightarrow \pi \ell \nu$,}
  \plb{644}{2007}{38}
  [arXiv:hep-ph/0611108].
  %%CITATION = PHLTA,B644,38;%%

%\cite{Vandewater_lattice_2008}
\bibitem{Vandewater_lattice_2008}
  R. Van~de~Water (for the Fermilab Lattice and MILC collaborations), talk at Lattice 2008 (Williamsburg, Virginia, USA, 14-19 July 2008).

%\cite{Bailey:2008wp}
\bibitem{Bailey:2008wp}
  J.~Bailey {\it et al.},
  %{\it The B $\rightarrow \pi \ell \nu$ semileptonic form factor from three-flavor lattice QCD: A model-independent determination of $|$V(ub)$|$,}
  arXiv:0811.3640 [hep-lat].
  %%CITATION = ARXIV:0811.3640;%%

%\cite{Bowler:1999xn}
\bibitem{Bowler:1999xn}
  K.~C.~Bowler {\it et al.}  [UKQCD Collaboration],
  %{\it Improved B $\rightarrow \pi \ell \nu_{\ell}$ form factors from the lattice,}
  \plb{486}{2000}{111}
  [arXiv:hep-lat/9911011].
  %%CITATION = PHLTA,B486,111;%%

%\cite{Isgur:1990kf}
\bibitem{Isgur:1990kf}
  N.~Isgur and M.~B.~Wise,
  %{\it Relationship between form-factors in semileptonic $\bar{B}$ and D decays and exclusive rare $\bar{B}$ meson decays,}
  \prd{\bf 42}{1990}{2388}.
  %%CITATION = PHRVA,D42,2388;%%

%\cite{Neubert:1993za}
\bibitem{Neubert:1993za}
  M.~Neubert,
  %{\it Short distance expansion of heavy - light currents at order 1/$m_Q$,}
  \prd{49}{1994}{1542}
  [arXiv:hep-ph/9308369].
  %%CITATION = PHRVA,D49,1542;%%

%\cite{Aoki:1999yr}
\bibitem{Aoki:1999yr}
  S.~Aoki {\it et al.}  [CP-PACS Collaboration],
  %{\it Quenched Light Hadron Spectrum,}
  \prl{84}{2000}{238}
  [arXiv:hep-lat/9904012].
  %%CITATION = PRLTA,84,238;%%

%\cite{Aubin:2007mc}
\bibitem{Aubin:2007mc}
  C.~Aubin and C.~Bernard,
  %{\it Heavy-Light Semileptonic Decays in Staggered Chiral  Perturbation Theory,}
  \prd{76}{2007}{014002}
  [arXiv:0704.0795 [hep-lat]].
  %%CITATION = PHRVA,D76,014002;%%

%\cite{Arndt:2004bg}
\bibitem{Arndt:2004bg}
  D.~Arndt and C.~J.~D.~Lin,
  %{\it Heavy meson chiral perturbation theory in finite volume,}
  \prd{70}{2004}{014503}
  [arXiv:hep-lat/0403012].
  %%CITATION = PHRVA,D70,014503;%%

%\cite{Bernard:2001av}
\bibitem{Bernard:2001av}
  C.~W.~Bernard {\it et al.},
  %{\it The QCD spectrum with three quark flavors,}
  \prd{64}{2001}{054506}
  [arXiv:hep-lat/0104002].
  %%CITATION = PHRVA,D64,054506;%%

%\cite{Okamoto:2004xg}
\bibitem{Okamoto:2004xg}
  M.~Okamoto {\it et al.},
  %{\it Semileptonic D $\rightarrow \pi / K$ and B $\rightarrow \pi / D$ decays in 2+1 flavor lattice QCD,}
  \npps{140}{2005}{461}
  [arXiv:hep-lat/0409116].
  %%CITATION = NUPHZ,140,461;%%
%\cite{ElKhadra:2001rv}
\bibitem{ElKhadra:2001rv}
  A.~X.~El-Khadra, A.~S.~Kronfeld, P.~B.~Mackenzie, S.~M.~Ryan and J.~N.~Simone,
  %``The semileptonic decays B --> pi l nu and D --> pi l nu from lattice
  %QCD,''
  Phys.\ Rev.\  D {\bf 64} (2001) 014502
  [arXiv:hep-ph/0101023].
  %%CITATION = PHRVA,D64,014502;%%

%\cite{Khodjamirian:2000ds}
\bibitem{Khodjamirian:2000ds}
  A.~Khodjamirian, R.~R\"uckl, S.~Weinzierl, C.~W.~Winhart and O.~I.~Yakovlev,
  %{\it Predictions on B $\rightarrow \pi \bar{\ell} \nu_\ell$, D $\rightarrow \pi \bar{\ell} \nu_\ell$ and D $\rightarrow K \bar{\ell} \nu_\ell$ from QCD light-cone sum rules,}
  \prd{62}{2000}{114002}
  [arXiv:hep-ph/0001297].
  %%CITATION = PHRVA,D62,114002;%%

%\cite{Ball:2006yd}
\bibitem{Ball:2006yd}
  P.~Ball,
 %{\it Testing QCD sum rules on the light-cone in D $\rightarrow (\pi, K) \ell \nu$ decays,}
  \plb{641}{2006}{50}
  [arXiv:hep-ph/0608116].
  %%CITATION = PHLTA,B641,50;%%

%\cite{Horgan:2009ti}
\bibitem{Horgan:2009ti}
  R.~R.~Horgan {\it et al.},
  %``Moving NRQCD for heavy-to-light form factors on the lattice,''
  arXiv:0906.0945 [hep-lat].
  %%CITATION = ARXIV:0906.0945;%%


%\cite{Aubert:2006px}
\bibitem{Aubert:2006px}
  B.~Aubert {\it et al.}  [BABAR Collaboration],
  %``Measurement of the $B^0 \to \pi^{-} \ell^{+} \nu$ form-factor shape and
  %branching fraction, and determination of $|V_{ub}|$ with a loose neutrino
  %reconstruction technique,''
  \prl{98}{2007}{091801} 
  %Phys.\ Rev.\ Lett.\  {\bf 98} (2007) 091801
  [arXiv:hep-ex/0612020].
  %%CITATION = PRLTA,98,091801;%%


%\cite{Flynn:2007ii}
\bibitem{Flynn:2007ii}
  J.~M.~Flynn and J.~Nieves,
  %``|Vub| from exclusive semileptonic B to pi decays revisited,''
  \prd{76}{2007}{031302}
  %Phys.\ Rev.\  D {\bf 76} (2007) 031302
  [arXiv:0705.3553 [hep-ph]].
  %%CITATION = PHRVA,D76,031302;%%



\end{thebibliography}
\end{document}